\documentclass[preprint]{ptephy_v1}
\preprintnumber{} 

\def\b8{$^8{\rm B}$}
\def\be7{$^7{\rm Be}$}

\newcommand{\nue}       {{\nu}_{\rm e}}

\newcommand{\numu}      {\nu_{\rm \mu}}

\newcommand{\nutau}     {{\nu}_{\rm \tau}}

\newcommand{\sstt}      {\sin^2 2\theta}

\newcommand{\dms}       {\Delta m^2}

\begin{document}

\title{History of Solar Neutrino Observations}


\author{Masayuki Nakahata \\
\small
  Kamioka Observatory, Institute for Cosmic Ray Research, University of Tokyo, \\
  456 Higashi-Mozumi, Kamioka-cho, Hida-shi, Gifu 506-1205, Japan\\
  Kavli Institute for the Physics and Mathematics of the Universe (WPI), \\
  The University of Tokyo Institutes for Advanced Study, University of Tokyo,
  Kashiwa, Chiba 277-8583, Japan
  \email{nakahata@icrr.u-tokyo.ac.jp}}
\normalsize


\begin{abstract}%
The first solar neutrino experiment led by Raymond Davis Jr. showed a deficit of neutrinos
relative to the solar model prediction, referred to as the
``solar neutrino problem'' since the 1970s. The Kamiokande experiment led by Masatoshi Koshiba successfully observed
solar neutrinos, as first reported in 1989.
The observed flux of solar neutrinos was almost half the prediction and confirmed
the solar neutrino problem.
This problem was not resolved for some time due to possible uncertainties
in the solar model.
In 2001, it was discovered that the solar neutrino problem is due to neutrino oscillations
by comparing the Super-Kamiokande and Sudbury Neutrino Observatory results, which was
the first model-independent comparison.
Detailed studies of solar neutrino oscillations have since been performed, and
the results of solar neutrino experiments are consistent with solar model
predictions when the effect of neutrino oscillations are taken into account.
In this article, the history of solar neutrino observations is reviewed
with the contributions of Kamiokande and Super-Kamiokande detailed.
\end{abstract}

\subjectindex{C43, F21}

\maketitle

\section{Introduction}
The main energy source of the sun is thermonuclear reactions inside the core.
Considerable amounts of electron-type neutrinos ($\nue$s) are produced 
through nuclear fusion reactions, and solar neutrino experiments provide
direct surveys of the deep interior of the sun.
Prof. R. Davis established the Homestake experiment to identify the main fusion
reactions in the 1960s.
However, the observed flux was much smaller than that predicted by
the standard solar model, and this was referred to as the ``solar neutrino problem''.
The Homestake experiment used a radiochemical method that collected
argon atoms produced by neutrino reactions.
Because of the unconventional technique of the experiment, it did not 
convince people whether the solar neutrino problem
could be attributed to properties of neutrinos themselves or errors in the 
standard solar model.

The Kamiokande experiment was undertaken in 1983 under the 
leadership of Prof. Masatoshi Koshiba.
The original purpose of Kamiokande was to search for proton decay.
Proton decays were not observed and Prof. Koshiba proposed
upgrading the detector for solar neutrino measurements a few months after
data collection was started.
In 1988, the Kamiokande experiment succeeded in observing solar neutrinos.
This was the first measurement with a real-time detection method.
The observed solar neutrino flux was about 50 \% of the prediction and
confirmed the solar neutrino problem.
Although the Homestake and Kamiokande experiments observed a deficit in the
solar neutrino flux, they could not determine the cause of the discrepancy because of
large uncertainties in the model predictions.

In the early 1990s, gallium experiments (SAGE and GALLEX) were
started to measure low-energy solar neutrinos.
These experiments also observed a flux smaller than the prediction,
enhancing the possibility of neutrino oscillations as the solution
of the solar neutrino problem.

In 1996, Super-Kamiokande, which had 30 times more fiducial volume
than Kamiokande, started collecting data. It detected about 22,400 
solar neutrino events by 2001 and the $^8$B solar neutrino flux was measured 
with an accuracy of 3\% using neutrino--electron scattering. 
In 2001, the SNO group announced a $^8$B flux measurement using charged-current
neutrino--deuteron interactions, and comparisons between the
Super-Kamiokande and SNO results gave direct evidence for 
model-independent solar neutrino oscillations.
The evidence was further strengthened by neutral-current measurements from SNO.
In 2002, combining the results of the solar neutrino experiments, global
analyses showed that the most suitable oscillation parameter is 
the large mixing angle (LMA; a solution with
a mass squared difference $\Delta m^2_{21} (=m_2^2 -m_1^2)$
of $10^{-5}-10^{-4}$ eV$^2$ and
a mixing angle ($\theta$) of $\sin^2(2\theta)=0.5-1$)\footnote{Details of the
oscillation parameters will be described in Sections \ref{SECPRESK} and \ref{SECOSC}.}.
In 2008, the Borexino experiment measured the flux of $^7$Be solar neutrinos
and further confirmed the existence of neutrino oscillations.

In this article, results from solar neutrino experiments are reviewed with
detailed descriptions of Kamiokande and Super-Kamiokande.
In Section 2, the standard solar model and its neutrino flux predictions
are described.
The results of the Homestake experiment are described in Section 3 and the
observations from Kamiokande are described in Section 4.
In Section 5, the status of understanding of the solar neutrino problem just
before the start of Super-Kamiokande and SNO is summarized, including results from
the SAGE and GALLEX experiments.
Solar neutrino measurements by Super-Kamiokande, SNO and Borexino are
described in Sections 6, 7 and 8, respectively.
Solar neutrino oscillations are discussed in Section \ref{SECOSC} and a conclusion
and future prospects are given in Section \ref{SECCONC}.

\section{Standard Solar Model} 

It is important to develop precise standard solar models in order
to discuss neutrino oscillations.
In the standard solar model (SSM)\cite{SOLBP2001,SOLBP2004,SSM2017},
the time evolution of the temperature and
pressure at each position in the sun is solved using equations of hydrostatic 
equilibrium, mass continuity, energy conservation and energy transport by
radiation or convection.
The boundary conditions for solving the model are the mass, radius, 
age and luminosity of the sun at present.
The input parameters for the SSM include nuclear fusion cross sections, the initial 
chemical composition of the sun (elements other than H and He) and 
the opacity to photons.
The SSM assumes that the current surface chemical composition reflects the
initial chemical composition, and photo-spectroscopic measurements of
the surface are used to estimate its chemical composition.

\begin{figure}
\begin{center}
\includegraphics[width=7cm]{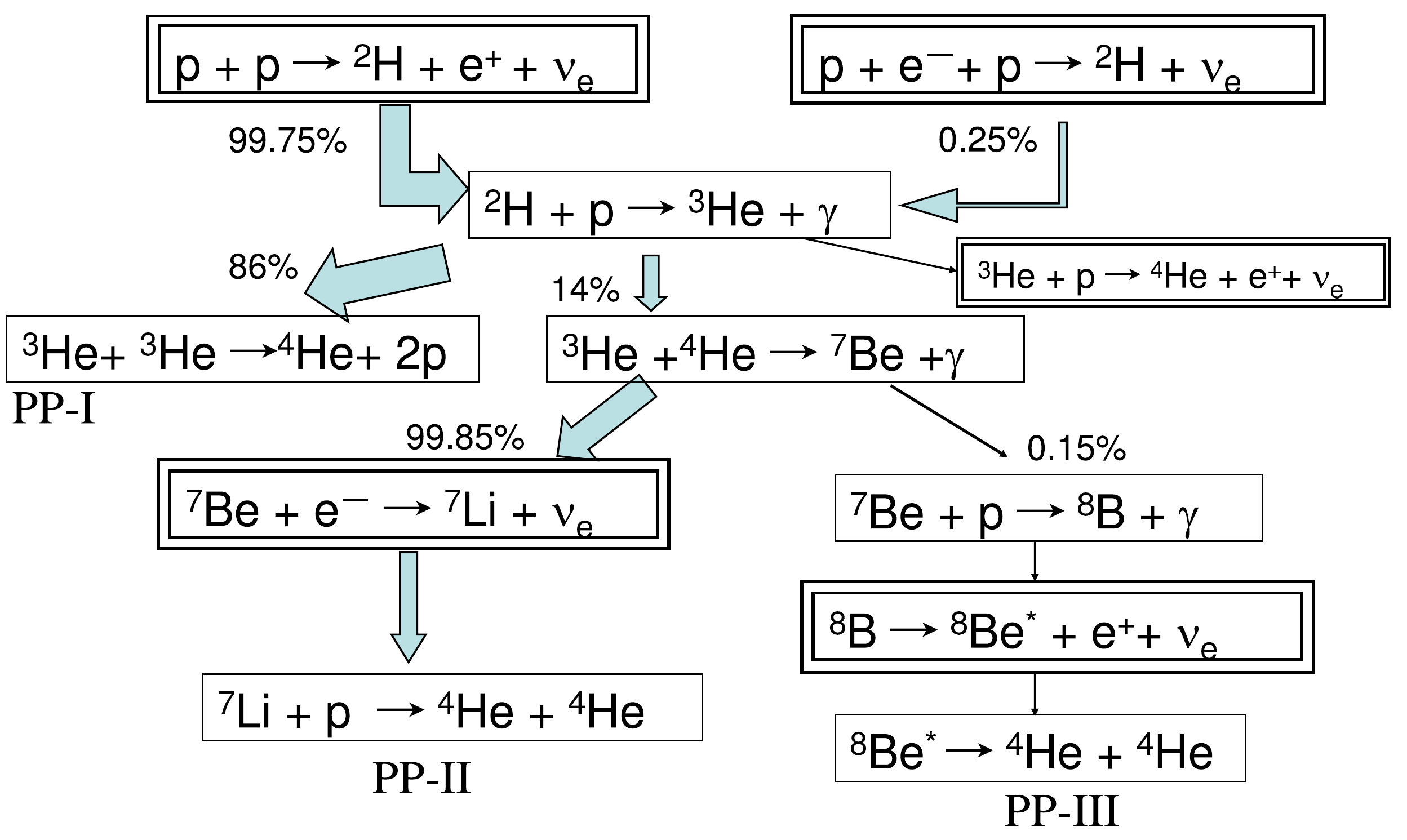}
\includegraphics[width=7cm]{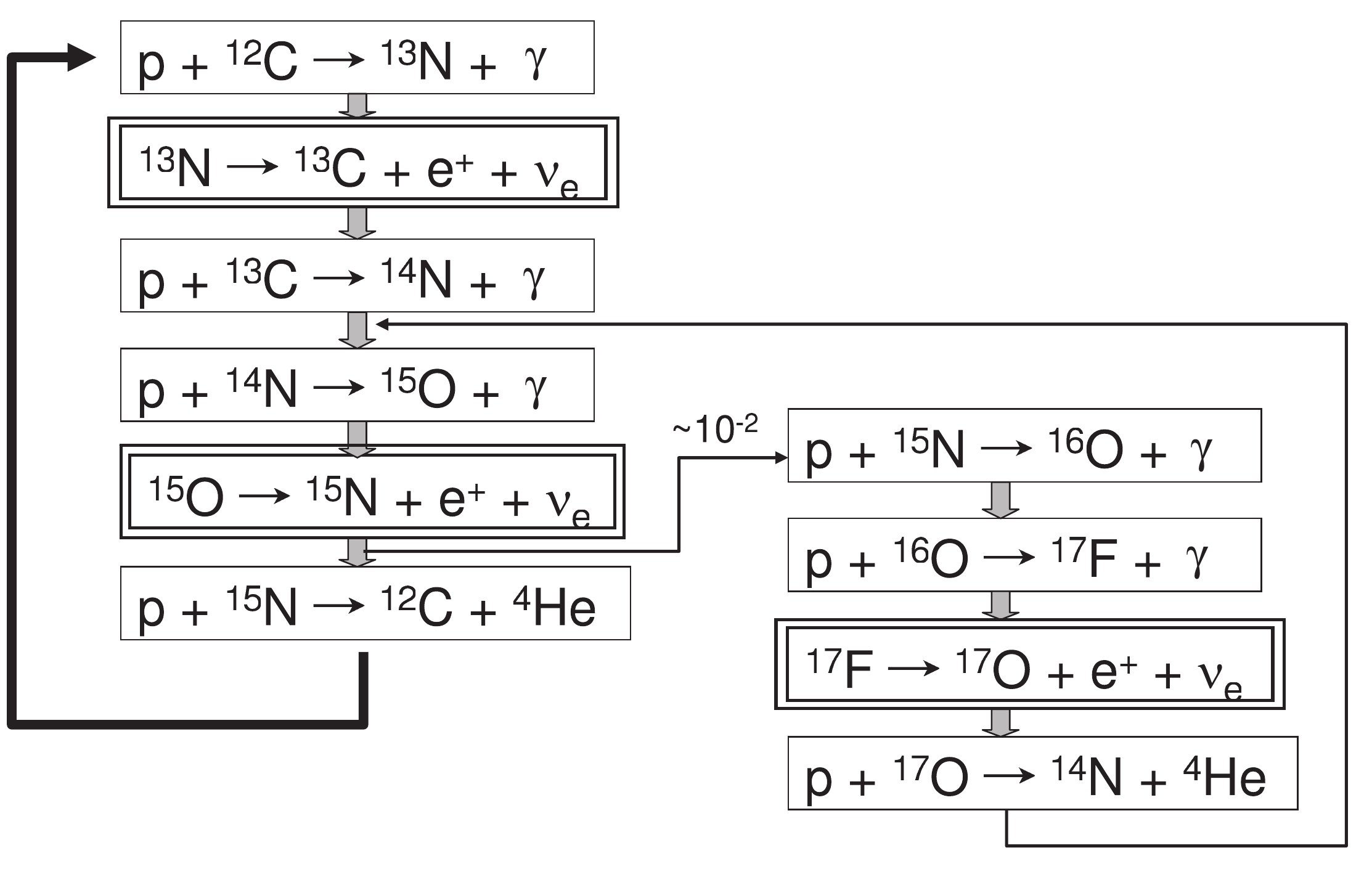}
\caption{$pp$ chain and CNO cycle reactions}
\label{sol-ssmchain}
\end{center}
\end{figure}

The SSM predicts that 99\% of the energy production
in the sun is due to the $pp$ nuclear reaction chain and
the remaining 1\% is due to the CNO cycle, as shown in 
Fig.\ref{sol-ssmchain}.
In the figure, the reactions marked with double borders produce neutrinos,
and the neutrinos from these reactions are named depending on the reaction: 
$pp$, $^7{\rm Be}$, $^8{\rm B}$, $hep$, $^{13}{\rm N}$, $^{15}{\rm O}$ and
$^{17}{\rm F}$ neutrinos.
The fluxes of each neutrino type from the latest SSM\cite{SSM2017}
are shown in Table \ref{sol-ssmtable}.
\begin{table}
\begin{center}
\begin{tabular}{|l|l|l|}
\hline
source & \multicolumn{2}{|c|}{flux (/cm$^2$/s)}\\
\hline
 & GS98 & AGSS09met \\
\hline \hline
$pp$ & 5.98 $\times 10^{10} (1 \pm 0.006)$ & 6.03 $\times 10^{10} (1 \pm 0.005)$ \\
\hline
$^7{\rm Be}$ & 4.93$\times 10^9 (1 \pm 0.06)$ & 4.50$\times 10^9 (1 \pm 0.06)$ \\
\hline
$pep$ & 1.44$\times 10^8 (1 \pm 0.01)$ & 1.46$\times 10^8 (1 \pm 0.009)$  \\
\hline
$^8{\rm B}$ & 5.46$\times 10^6 (1 \pm 0.12)$ & 4.50$\times 10^6 (1 \pm 0.12)$ \\
\hline
$hep$ & 7.98$\times 10^3 (1 \pm 0.30)$ & 8.25$\times 10^3 (1 \pm 0.30)$  \\
\hline
$^{13}{\rm N}$ & 2.78$\times 10^8 (1 \pm 0.15)$ & 2.04$\times 10^8 (1 \pm 0.14)$ \\
\hline
$^{15}{\rm O}$ & 2.05$\times 10^8 (1\pm 0.17)$ & 1.44$\times 10^8 (1\pm 0.16)$ \\
\hline
$^{17}{\rm F}$ & 5.29$\times 10^6 (1\pm 0.20)$ & 3.26$\times 10^6 (1\pm 0.18)$ \\
\hline
\end{tabular}
\end{center}
\caption{Solar neutrino flux predicted by SSM\cite{SSM2017}.
The second and third columns show the flux predictions using the chemical
composition from GS98\cite{GS98} and AGSS09met\cite{AGSS09}, respectively.} \label{sol-ssmtable} 
\end{table}
The second and third columns in the table show the flux predictions using
chemical compositions obtained by GS98\cite{GS98} and AGSS09met\cite{AGSS09}, 
respectively.
GS98 is based on a one-dimensional model of the solar atmosphere 
that was released in 1998.
AGSS09met, released in 2009, is based on a three-dimensional model 
and uses the most up-to-date atomic and molecular data,
and should therefore be more reliable than the GS98-based 
solar model.
However, the GS98-based solar model can reproduce various observations
inside the sun, such as the sound speed profile, depth of the convective 
zone and the helium abundance, while the AGSS09met based solar model's 
predictions have large discrepancies with observations.
Therefore, the flux predictions of both are given here.

\begin{figure}
\begin{center}
\includegraphics[width=12cm]{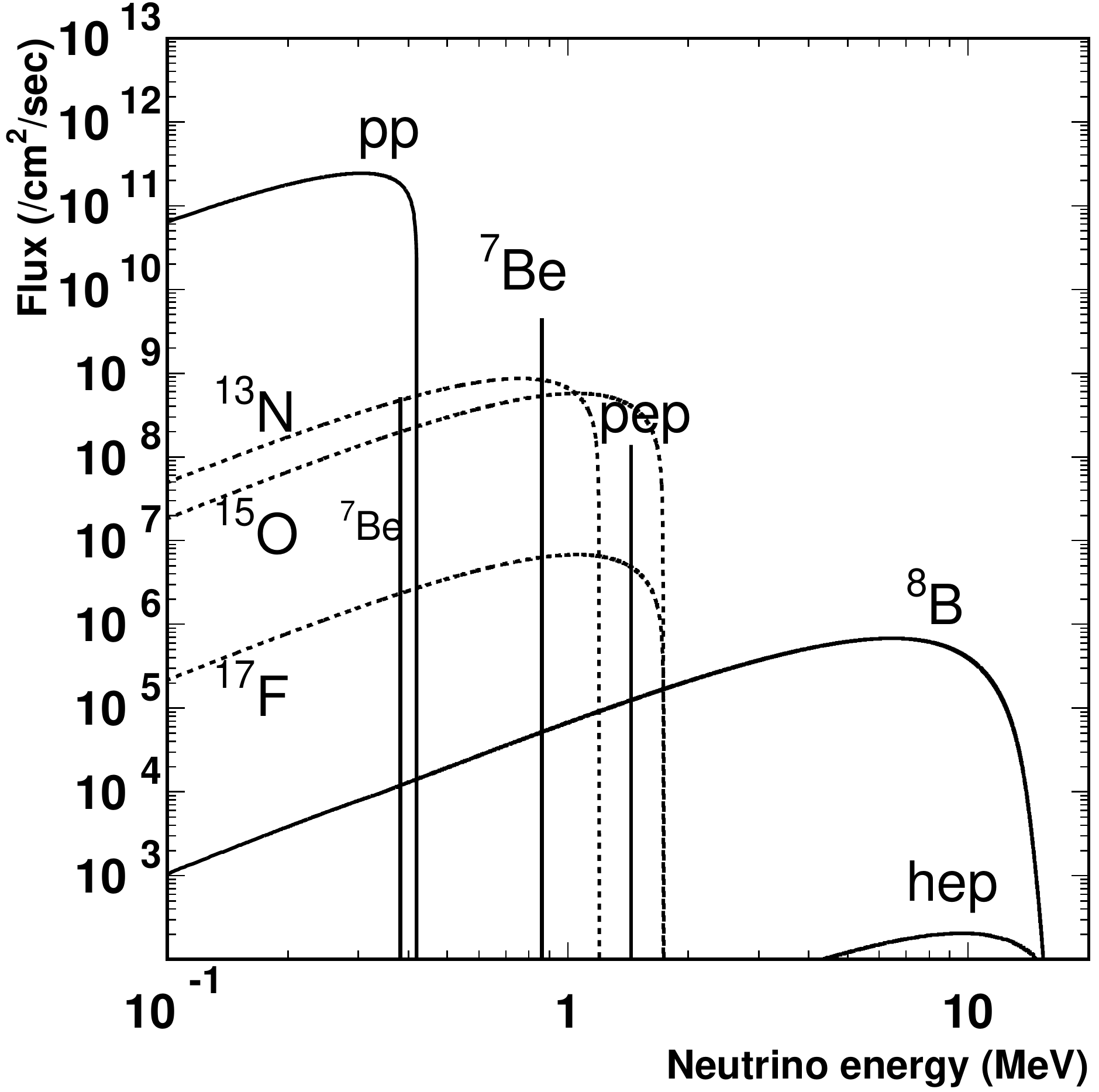}
\caption{Solar neutrino energy spectrum expected from the SSM\cite{SOLBP2001,SOLBP2004}.
The solid and dashed curves show neutrinos in the $pp$ chain and CNO cycle,
respectively.}
\label{sol-ssmspec}
\end{center}
\end{figure}

The energy spectrum of solar neutrinos predicted by the SSM is shown in 
Fig.\ref{sol-ssmspec}.
The most energetic neutrino is the $^8$B neutrino and it was the main 
neutrino source for the Homestake, Kamiokande, Super-Kamiokande and SNO 
experiments, though its intensity is only about 0.01\% of the
total solar neutrino flux.
The most abundant source is the $pp$ neutrino but its maximum energy is only
0.42 MeV. The gallium experiments were sensitive to $pp$ neutrinos.

\section{Homestake experiment} 

\begin{figure}
\begin{center}
\includegraphics[width=15cm]{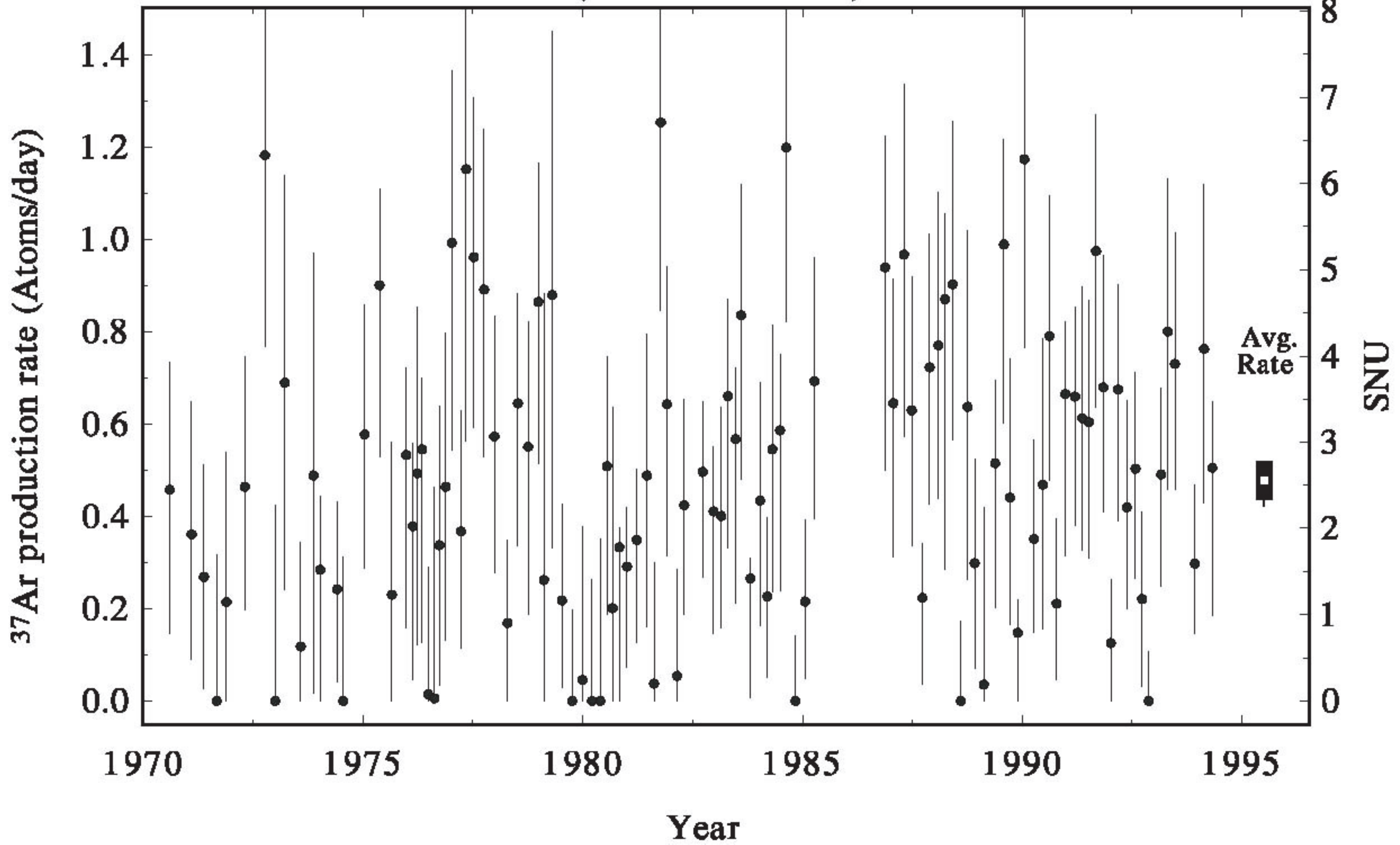}
\caption{Data from the Homestake experiment\cite{SOLDAVIS}. Each data point 
shows the $^{37}{\rm Ar}$ production rate measured in each collection cycle (run).
The scale of the right hand side is in SNU.
The time averaged rate is shown to the right of the main data.}
\label{sol-homestake}
\end{center}
\end{figure}

The Homestake experiment was located in the Homestake gold mine at a
depth of 1480 meters\cite{SOLDAVIS}.
The experiment was started around 1970 and data were obtained until 1994. 
The target for the solar neutrinos was $^{37}{\rm Cl}$ atoms in 615 tons of
${\rm C}_2{\rm Cl}_4$.
The neutrino energy threshold of the reaction 
$^{37}{\rm Cl}+\nu_e \rightarrow ^{37}{\rm Ar}+e^-$ is 0.814 MeV and it is
mainly sensitive to $^8$B neutrinos.
This radiochemical Cl-Ar method of
solar neutrino detection was proposed by B. Pontecorvo in 1946\cite{PONT1946}.
The expected event rate from the SSM\cite{SOLBP2004} 
was 8.5$\pm$1.8 SNU, where one SNU is
$10^{-36}$captures/atom/s. The contribution from each neutrino source is
6.6 SNU from $^8{\rm B}$ neutrinos, 1.2 SNU from $^7$Be neutrinos, 0.22 SNU
from $pep$ neutrinos and the remainder from CNO cycle neutrinos.
The produced $^{37}{\rm Ar}$ atoms were collected once every 60--120 days and
the decay of $^{37}{\rm Ar}$ was counted using a low background proportional counter.
Figure \ref{sol-homestake} shows the observed production rate of $^{37}{\rm Ar}$ in
each collection cycle\cite{SOLDAVIS}.
The average event rate observed by the Homestake experiment was
\begin{eqnarray}
    \phi(Homestake) = 2.56 \pm 0.16(stat.) \pm 0.16(sys.) ~~ {\rm SNU}
\nonumber
\end{eqnarray}
The observed event rate was only about 30\% of the SSM prediction,
leading to the so-called solar neutrino problem.

\section{Kamiokande} 

\begin{figure}
\begin{center}
\includegraphics[width=12cm]{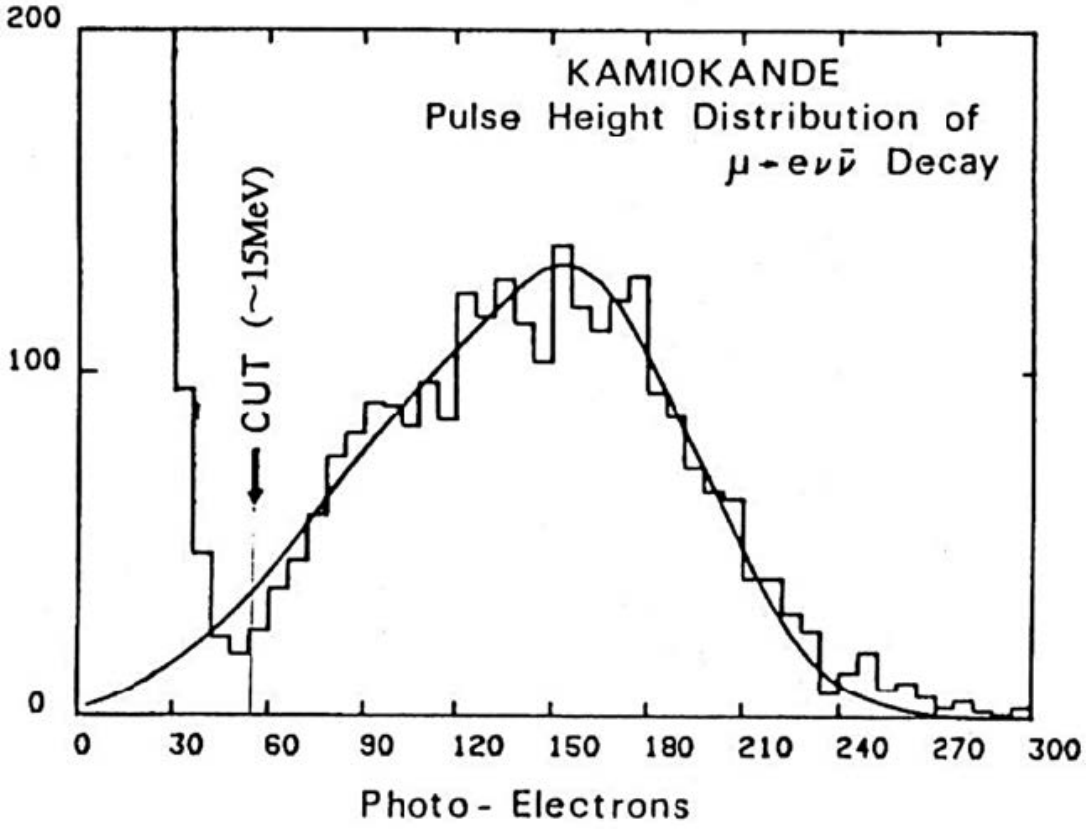}
\caption{Pulse height distribution of $\mu \rightarrow e$
  signals of stopping cosmic ray muons at the early stage of
  Kamiokande.}
\label{muedecay}
\end{center}
\end{figure}

The Kamiokande detector was constructed in 1983 to search for proton decay.
The detector had a 2340 ton water volume, located 1000 meters underground in
the Kamioka Mine in Japan.
The water volume was viewed by 1,000 20-inch diameter
photomultipliers (PMTs) mounted on a 1-m grid on the inner surface. 
Data collection was started on July 6, 1983 and tens of
proton decay events within a few months were expected if the
original idea of the grand unified theory\cite{GGGUT} is correct.
However, no proton decay events were found.
Since the primary purpose of the detector was to observe events
with $\sim$1 GeV total energy, low-energy events such as solar
neutrinos were not triggered in the first stage of the
Kamiokande experiment
as the electronics at that time read out only the integrated charge
information for each PMT.
The total sum signal, which is the sum of analog signals from all PMTs,
was used to trigger the readout electronics, and the trigger energy
threshold was about 30 MeV for electrons, which is much higher than
the energy of solar neutrino signals.
The total sum signal was recorded by a
transient digitizer (R7912, Tektronix), which records a
digitized oscilloscope image in order to detect
$\mu \rightarrow e$ decay signals from stopping muons.
There were several hundred cosmic ray stopping muons per day in the Kamiokande detector.
Figure \ref{muedecay} shows the pulse height distribution of $\mu \rightarrow e$
signals.
The distribution showed that the Kamioka detector had the potential to
detect low-energy neutrinos down to $\sim$10 MeV, and
Prof. Koshiba proposed upgrading the detector for \b8 solar
neutrino measurements in the fall of 1983.

In order to detect solar neutrinos, two major upgrades were required.
First, an anti-counter system had to be constructed to reduce
environmental background signals such as gamma rays from radioactivity in the
surrounding rock.
Second, a new set of electronics to read out the timing and charge
information of each individual PMT was necessary.
Precise determination of the position where an event occurred,
called the vertex position, is crucial for solar neutrino
measurements.
The gamma rays from the rock which remained after passing through
the anti-counter tended to have a vertex position near the wall of the
detector.
High-energy cosmic rays produce hadronic cascade showers by spallation
and generate short-life radioactive nuclei.
The event rate of the beta decays of those nuclei was more than one order
of magnitude higher than the expected rate of \b8 solar neutrino
signals.
Those spallation products were produced along the track of the originating
cosmic ray muons and could be removed using the vertex information.

\begin{figure}
\begin{center}
\includegraphics[width=14cm]{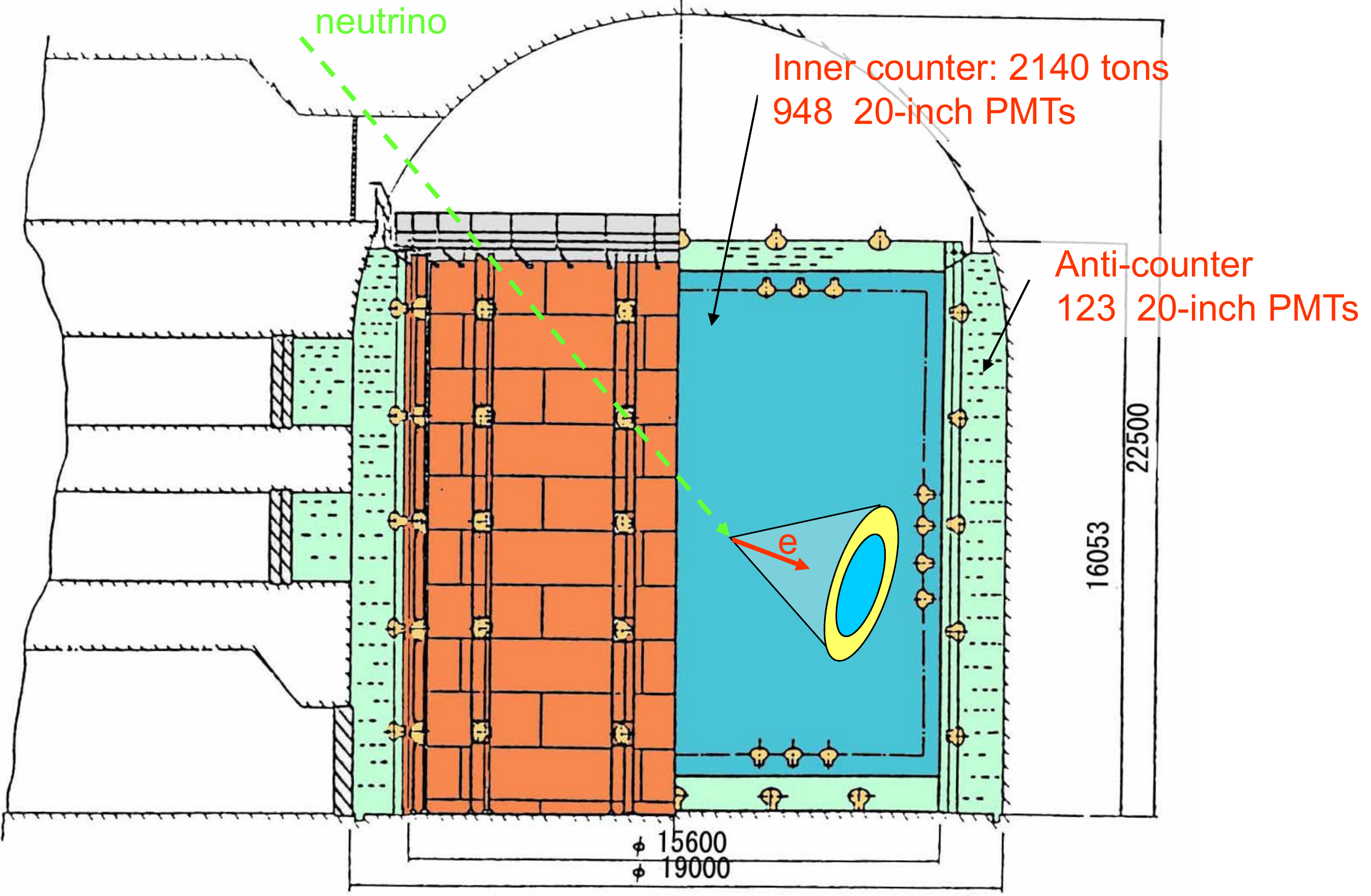}
\caption{Schematic view of the Kamiokande detector after the upgrade.}
\label{Kamiokande}
\end{center}
\end{figure}

The anti-counter was constructed from September 1984 through March 1985.
123 20-inch PMTs were mounted in the anti-counter to
work as an active veto.
The bottom PMT plane of the inner-counter was lifted by 1.2 m 
to allow PMTs to be inserted for the bottom anti-counter.
The PMTs for the top anti-counter were mounted 0.8 m above the top
inner-counter.
Additional water was added to the tank to submerge these PMTs. 
The Kamiokande tank was constructed in a newly excavated cavern,
and rubber-asphalt was sprayed on the surface of the cavern to make it
water-tight so that water could be filled between the cavern wall and the Kamiokande tank.
PMTs were mounted in this newly prepared layer, which functioned as the barrel part of
the anti-counter.
A schematic view of the Kamiokande detector after the upgrade is shown in
Fig.\ref{Kamiokande}.

Prof. Koshiba gave a talk on the possibility of detecting \b8
solar neutrinos in Kamiokande at ICOBAN'84 (International Conference
On BAryon Non-conservation, Park City, Utah, 1984)
and asked the participants to collaborate on the electronics upgrade.
Soon after this workshop, the Univ. Pennsylvania group led by Prof. A.K. Mann
expressed interest in producing the new electronics.
The electronics system, included timing and charge readouts and a trigger
system using number-of-hit PMTs, was installed in the fall of 1985.

\begin{figure}
\begin{center}
\includegraphics[width=8cm]{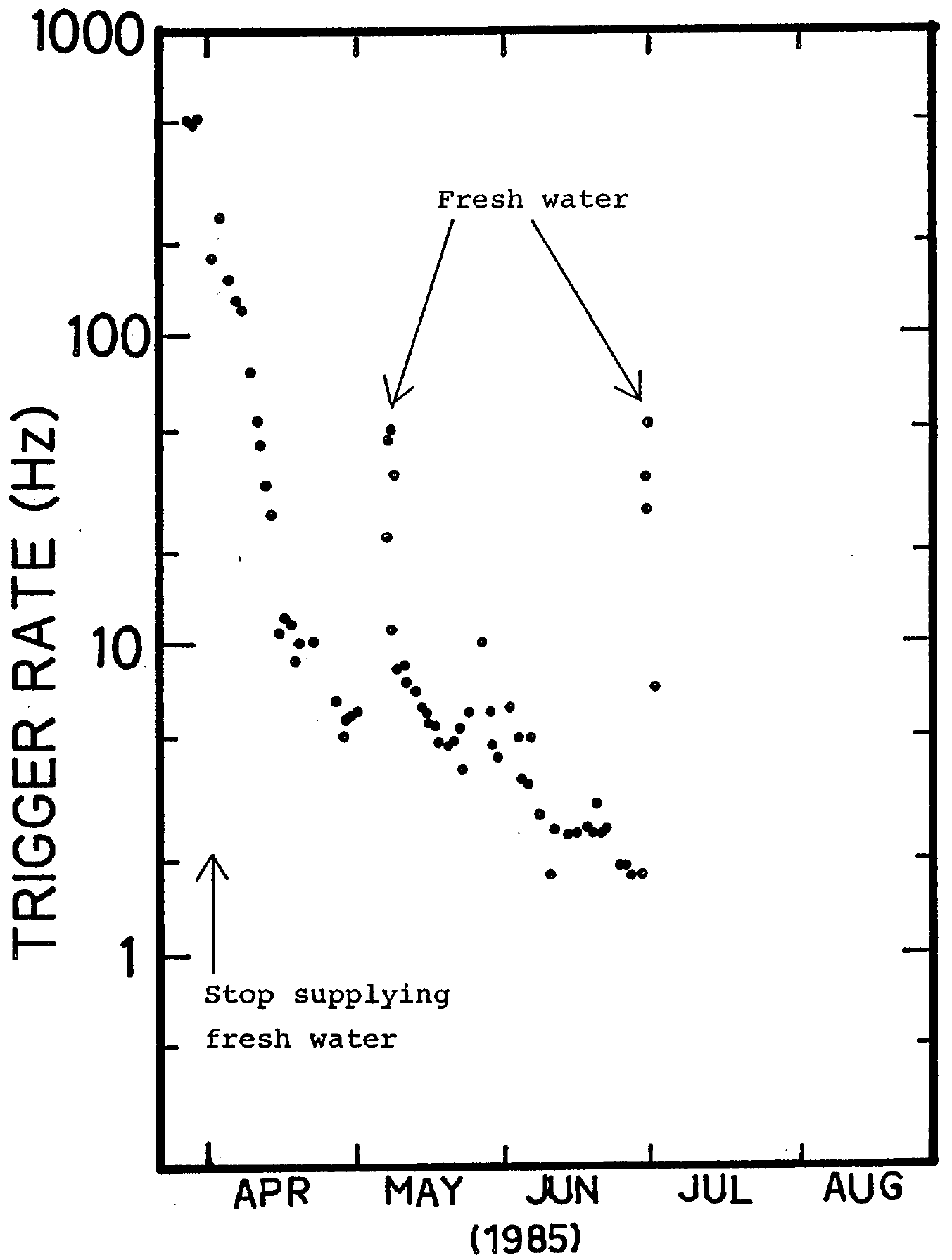}
\caption{Change in trigger rate after changing to
recirculated water without supplying fresh water.}
\label{kamtrigrate}
\end{center}
\end{figure}

A further consideration was radon in the tank water, which is the most significant source of background noise in solar neutrino measurements.
The concentration of radon, especially $^{222}{\rm Rn}$, was two or more orders of
magnitude higher in the air and water in the mine than in the usual environment
outside the mine.
A daughter of $^{222}{\rm Rn}$, $^{214}{\rm Bi}$, emits beta rays with an
end-point energy of 3.27 MeV, which mimics a solar neutrino signal.
When Kamiokande started collecting data after the installation of anti-counters,
fresh water was being constantly supplied to the tank.
The trigger rate in the lower energy threshold mode was several hundred events
per second, while the goal was at most only about one solar neutrino event per day. 
Figure \ref{kamtrigrate} shows the change in trigger rate after a modification was made to
recirculate water rather than supplying fresh water.
The decrease in the trigger rate was consistent with the decay of
$^{222}{\rm Rn}$ with a half life of 3.8 days.
Various efforts were made to further reduce radon in the tank water in 1986, including air-tightening of buffer tanks in the water
purification system.
The detector was almost ready for solar neutrino measurements in
early 1987.

Prof. Koshiba noted that the following three features are necessary to
perform ``neutrino astronomy'':
\begin{itemize}
\item Directionality
\item Real-time measurement
\item Energy spectrum measurement
\end{itemize}
In Kamiokande, solar neutrino signals were observed based on the Cherenkov radiation
of recoil electrons from neutrino--electron scattering.
Since the energy of \b8 solar neutrino is much larger than the
mass of an electron, the direction of a scattered electron is strongly
correlated with the direction from the sun to the earth.
In addition, since Kamiokande observed images of the Cherenkov ring pattern,
the direction of the electron could be observed.
Kamiokande was a real-time detector and events were observed at the time
when they happened.
The energy spectrum of the neutrinos could be deduced from the spectrum of
the scattered electrons.
Thus, the observations at Kamiokande satisfied the criteria for
neutrino astronomy.

\begin{figure}
\begin{center}
\includegraphics[width=8cm]{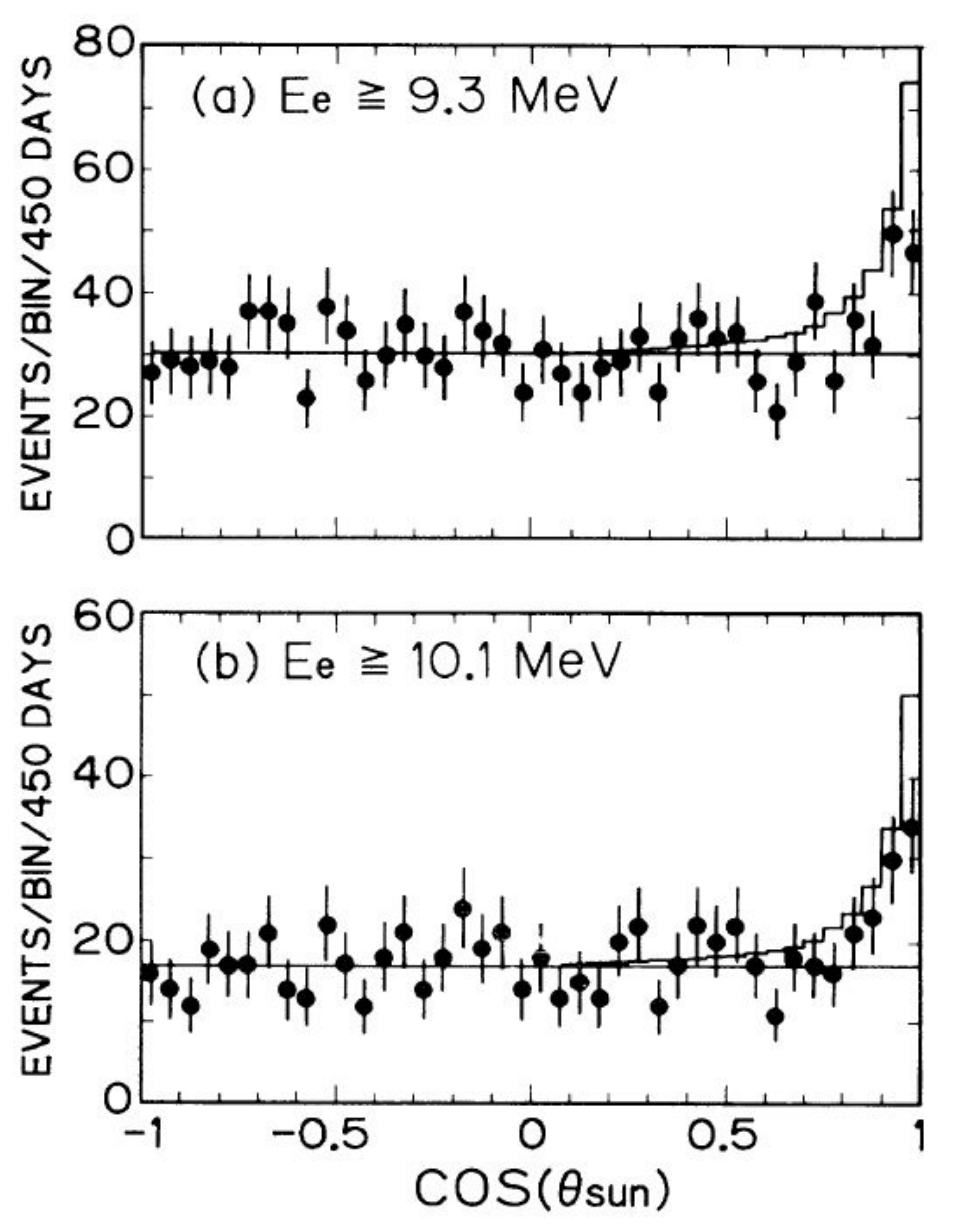}
\caption{Angular distribution to the direction of the sun
in Kamiokande\cite{SOLKAMIOKANDE1}. The plot shows data from the first 
450 days of data taken from January 1987 to May 1988. The solid histogram
shows the prediction from the SSM.}
\label{kam450d}
\end{center}
\end{figure}

Figure \ref{kam450d} shows the angular distribution to the sun for the
events that passed the criteria for solar neutrino event selection.
The plot was obtained with initial 450-day data taken from January 1987 to 
May 1988\cite{SOLKAMIOKANDE1}.
A clear excess of events was observed in the direction of the sun
but the observed rate was about 50\% of the prediction from the
SSM (solid histogram in the figure).
This observation confirmed the solar neutrino problem.

\begin{figure}
\begin{center}
\includegraphics[width=12cm]{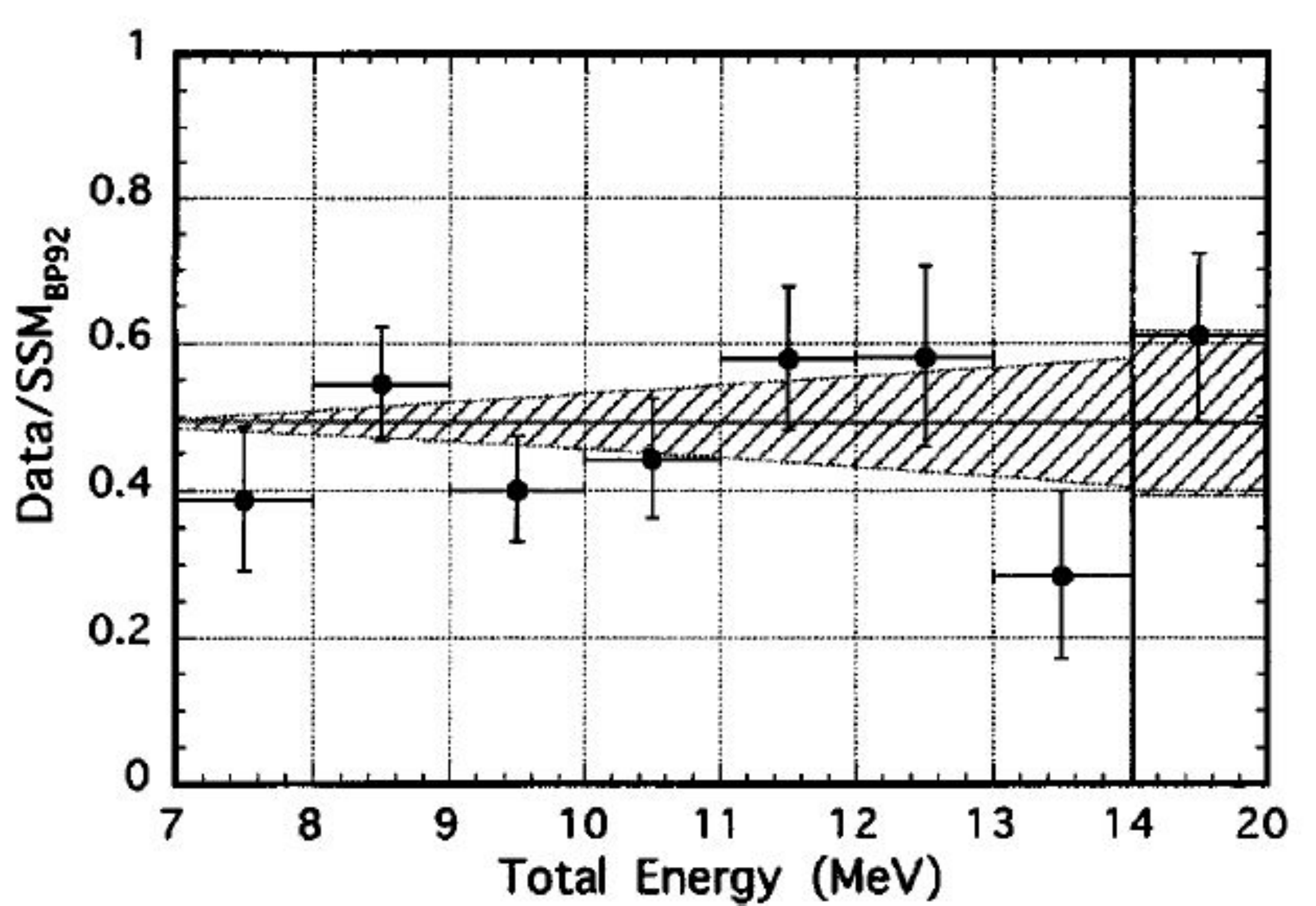}
\caption{Energy spectrum of recoil electrons normalized by the predicted 
spectrum from the 2079-day dataset of Kamiokande.
The hatched area shows the range of systematic uncertainty.}
\label{kamspec}
\end{center}
\end{figure}

The Kamiokande detector observed $\sim$600 solar neutrino events by
February 1995\cite{SOLKAMIOKANDE2} and the obtained flux of $^8$B neutrino was
\begin{eqnarray}
 \phi(^8{\rm B})_{Kamiokande} &=& 2.80 \pm 0.19 (\mbox{stat.}) \pm 0.33 (\mbox{sys.})
\nonumber \\
    & & \times 10^6 /cm^2/s~. \nonumber
\end{eqnarray}
The observed flux was 48$\pm$3(stat.)$\pm$6(sys.) \% of the prediction
from the SSM.
The energy spectrum of recoil electrons normalized by the predicted spectrum
is shown in Fig.\ref{kamspec},
consistent with a flat spectrum, i.e. no hint of neutrino
oscillations was given in the spectrum.
In order to proceed to investigate possibility of neutrino
oscillations, high statistics measurements with Super-Kamiokande
was necessary.

\section{Status just before the start of Super-Kamiokande and SNO}\label{SECPRESK}

This section describes other experiments that have been performed and
how we enhanced our understanding of the solar neutrino
problem before the Super-Kamioakande/SNO era.

SAGE and GALLEX were radiochemical experiments using gallium targets
conducted since the early 1990s.
The SAGE experiment\cite{SAGE1994,SAGE2001,SAGE2009} was conducted 
at the Baksan Observatory and the GALLEX experiment 
\cite{GALLEX94,GALLEX95,GALLEX96}
(later changed to GNO\cite{SOLGNO})
was conducted at the Gran Sasso Laboratory.
The energy threshold of the gallium reaction 
(\mbox{$^{71}{\rm Ga}+\nu_e \rightarrow ^{71}{\rm Ge}+e^-$}) is 0.233 MeV and 
is mainly sensitive
to low-energy solar neutrinos.
The expected event rate from the SSM\cite{SOLBP2004} is 
131$^{+12}_{-10}$ SNU,
with contributions of
69.6 SNU from $pp$ neutrinos, 34.8 SNU from $^7$Be neutrinos, 13.9 SNU
from \b8 neutrinos, 2.9 SNU from $pep$ neutrinos and the remainder
from CNO cycle neutrinos.
SAGE used 50 tons of gallium in metallic form and GALLEX/GNO used
30 tons of gallium in a GaCl$_3 \cdot$HCl solution.
The lifetime of $^{71}$Ge is 16.5 days and a typical exposure time 
for one run was 28 days.
By the middle of 1990s, a deficit of solar
neutrinos had been observed at SAGE and GALLEX, where the predicted
event rate was mainly coming from robust sources such as
$pp$ and $^7$Be neutrinos.
The final results of the gallium experiments are shown here but the
conclusion had not been changed since the middle of the 1990s.
The average event rates observed by SAGE and GALLEX/GNO were
\begin{eqnarray}
    \phi(SAGE) = 65.4 ^{+3.1}_{-3.0} (stat.) ^{+2.6}_{-2.8}(sys.) ~{\rm SNU} \nonumber \\
    \phi(GALLEX) = 73.1 ^{+6.1}_{-6.0}(stat.) ^{+3.7}_{-4.1} (sys.) ~{\rm SNU} \nonumber \\
    \phi(GNO) = 62.9 ^{+5.5}_{-5.3}(stat.) ^{+2.5}_{-2.5} (sys.) ~{\rm SNU}~. \nonumber
\end{eqnarray}
Combining these results\cite{SAGE2009}, the flux measured by the gallium 
experiments was
\begin{eqnarray}
    \phi(gallium) = 66.1 \pm 3.1 ~~ {\rm SNU} \nonumber
\end{eqnarray}
The observed flux is 50\% of the expectation from the SSM\cite{SOLBP2004}.

\begin{figure}
\begin{center}
\includegraphics[width=8cm]{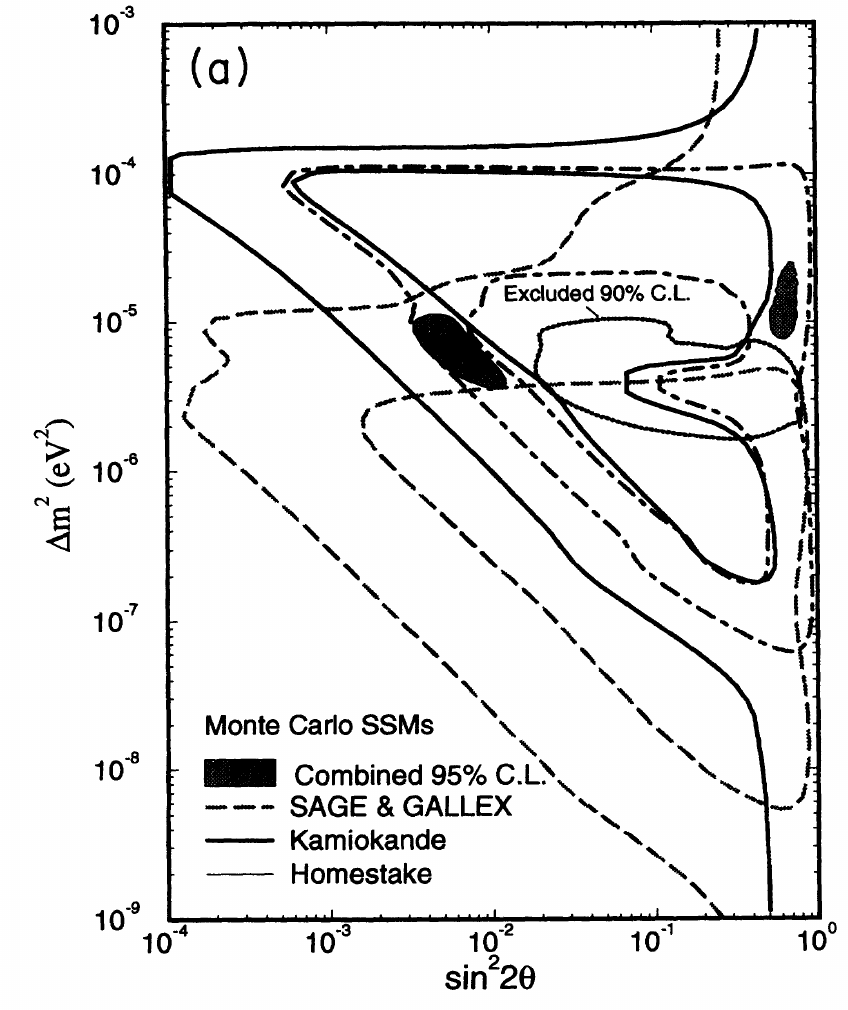}
\caption{Global analysis performed by Hata and Langacker in 1993\cite{HATALANG1993}.}
\label{HATALANG}
\end{center}
\end{figure}

Since it was difficult to attribute the deficits of
the neutrino event rates at the Homestake, Kamiokande and SAGE/GALLEX
experiments to potential
problems in the SSM, the possibility of solar neutrino oscillations was
extensively discussed in the mid 1990s.
Assuming two types of neutrinos, the relation between the mass eigenstates
of the two neutrinos ($\nu_1$ and $\nu_2$) and their interaction 
eigenstates($\nu_e$ and $\nu_X$(X=$\mu,\tau$))
is expressed as
\begin{eqnarray}
\left(
\begin{array}{c}
\nu_e \\
\nu_X \\
\end{array}
\right) &=&
\left(
\begin{array}{cc}
\cos \theta & \sin \theta \\
- \sin \theta & \cos \theta \\
\end{array}\right)
\left(
\begin{array}{c}
\nu_1 \\
\nu_2 \\
\end{array}
\right) ~,
\nonumber
\end{eqnarray}
where $\theta$ is the mixing angle.
Solving the time evolution of the neutrino wave function, the
probability that produced as electron-type neutrinos are observed as
eletron-type neutrinos is
\begin{eqnarray}
P(\nu_e \rightarrow \nu_e) & = & 
1 - \sin^2 2\theta \times \sin^2 \left( 1.27 \times \dms \frac{L}{E} \right)
~,
\nonumber
\end{eqnarray}
where $\dms$ is the mass squared difference ($m_2^2 -m_1^2$) in units of
eV$^2$, $L$ is the
neutrino travel length in meters, and $E$ is neutrino energy in MeV. 
If the argument of the last sine function, 
$1.27 \times \dms \frac{L}{E}$, is much 
larger than $2\pi$, it averages out and
the survival probability becomes :
\begin{eqnarray}
P(\nu_e \rightarrow \nu_e) & = & 1 - {1 \over 2}\sin^2 2\theta ~.
\nonumber
\end{eqnarray}
In the case of solar neutrino oscillation, the effect of matter
in the sun and the earth must be considered.
The matter effect was originally discussed by
L. Wolfensterin\cite{SOLMSWW} in 1978.
P. Langacker, J. P. Leveille and J. Sheiman
corrected a sign mistake in the Wolfensterin's paper in 1983\cite{LLS1983}.
Then in 1985, S. P. Mikheyev and A. Y. Smirnov pointed out that
the significant deficit of the event rate
in Homestake, especially much less than the half of the expectation,
can be explained by the matter effect\cite{SOLMSWMS}.
As you can see in the upper equation, the $P(\nu_e \rightarrow \nu_e)$
cannot be less than 0.5, but the matter effect can make it much
less than 0.5 if $\dms > 0$, i.e. $m_2 > m_1$.
Thus, the mass ordering of $m_2$ and $m_1$ was determined by
the solar neutrino observations.
In 1993, global analyses had been performed\cite{KRAPET1993,HATALANG1993} and
a typical contour plot of the oscillation parameters at that time
is shown in Fig.\ref{HATALANG}.
The two solutions shown in the figure are called the
SMA (small mixing angle) and LMA (large mixing angle) solutions for
$\sstt \sim 10^{-2}$ and $\sstt \sim 0.6$, respectively.

\begin{figure}
\begin{center}
\includegraphics[width=12cm]{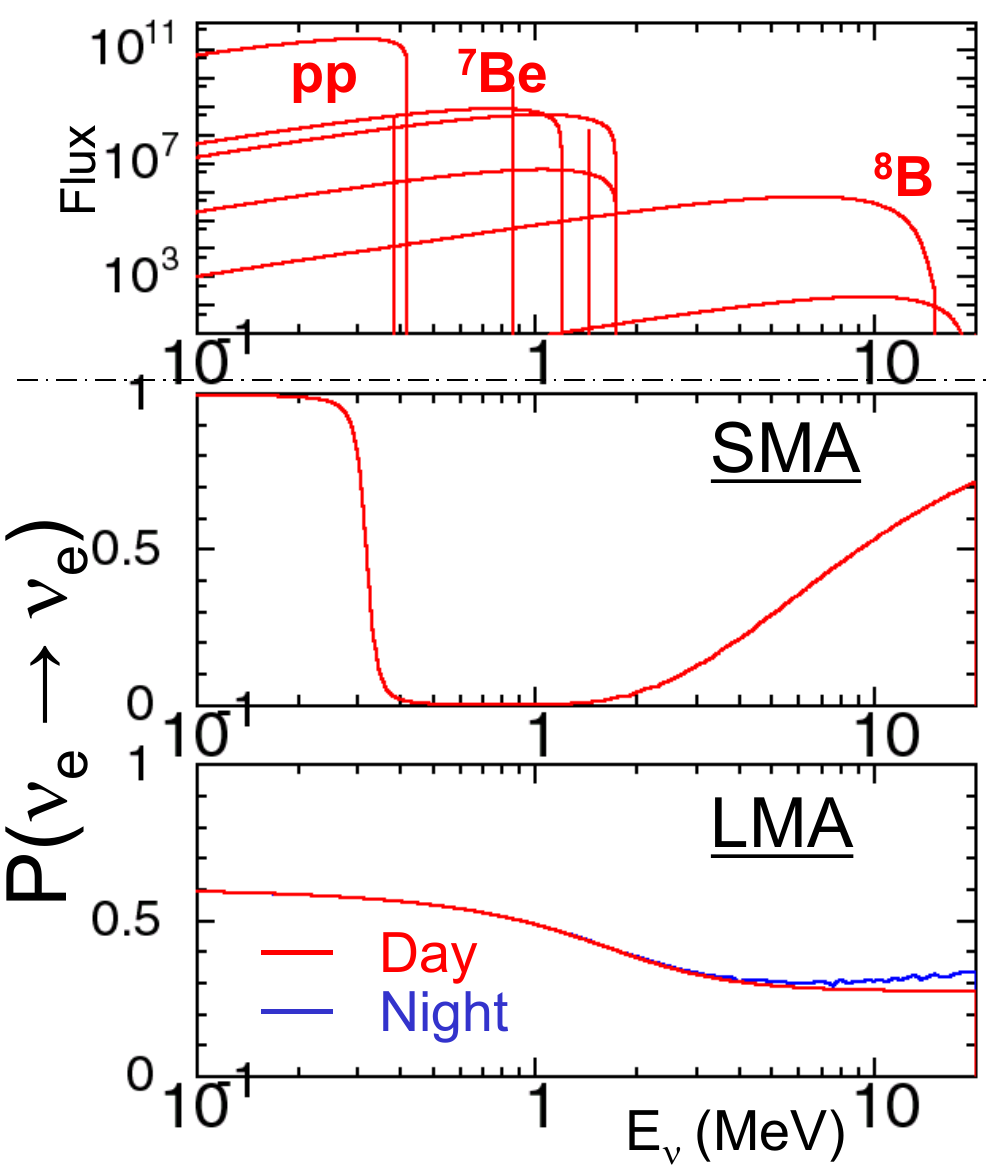}
\caption{Neutrino survival probability as a function of energy for typical
  oscillation parameters in each solution.
  The top plot is the predicted solar neutrino spectrum from the SSM.}
\label{OSCPROB}
\end{center}
\end{figure}

To obtain information about neutrino oscillations, it was necessary to
measure something that is independent of solar models.
Possibilities were the shape of the energy spectrum and the time
variation of solar neutrino event rates, such as the day/night difference.
Another possibility, which is more robust, was a
comparison between event rates measured with charged current
and neutral current interactions, which are sensitive to only $\nue$ and
all neutrino types, respectively.
Regarding the shape of the energy spectrum and the day/night difference,
Fig.\ref{OSCPROB} shows the neutrino survival probability as a function
of energy for each solution.

Super-Kamiokande aimed to measure the energy spectrum shape distortion,
which is expected for SMA, and the day/night difference, which is
expected for LMA with smaller $\dms$.
As will be described in Section \ref{SECOSC}, the first model-independent
discovery of neutrino oscillations was in a comparison between a charged
current measurement by SNO (i.e. $\nue$ flux measurement)
and an electron-scattering measurement by Super-Kamiokande (which had contributions from
$\numu$ and $\nutau$).

\section{Super-Kamiokande} 

\begin{figure}
\begin{center}
\includegraphics[width=12cm]{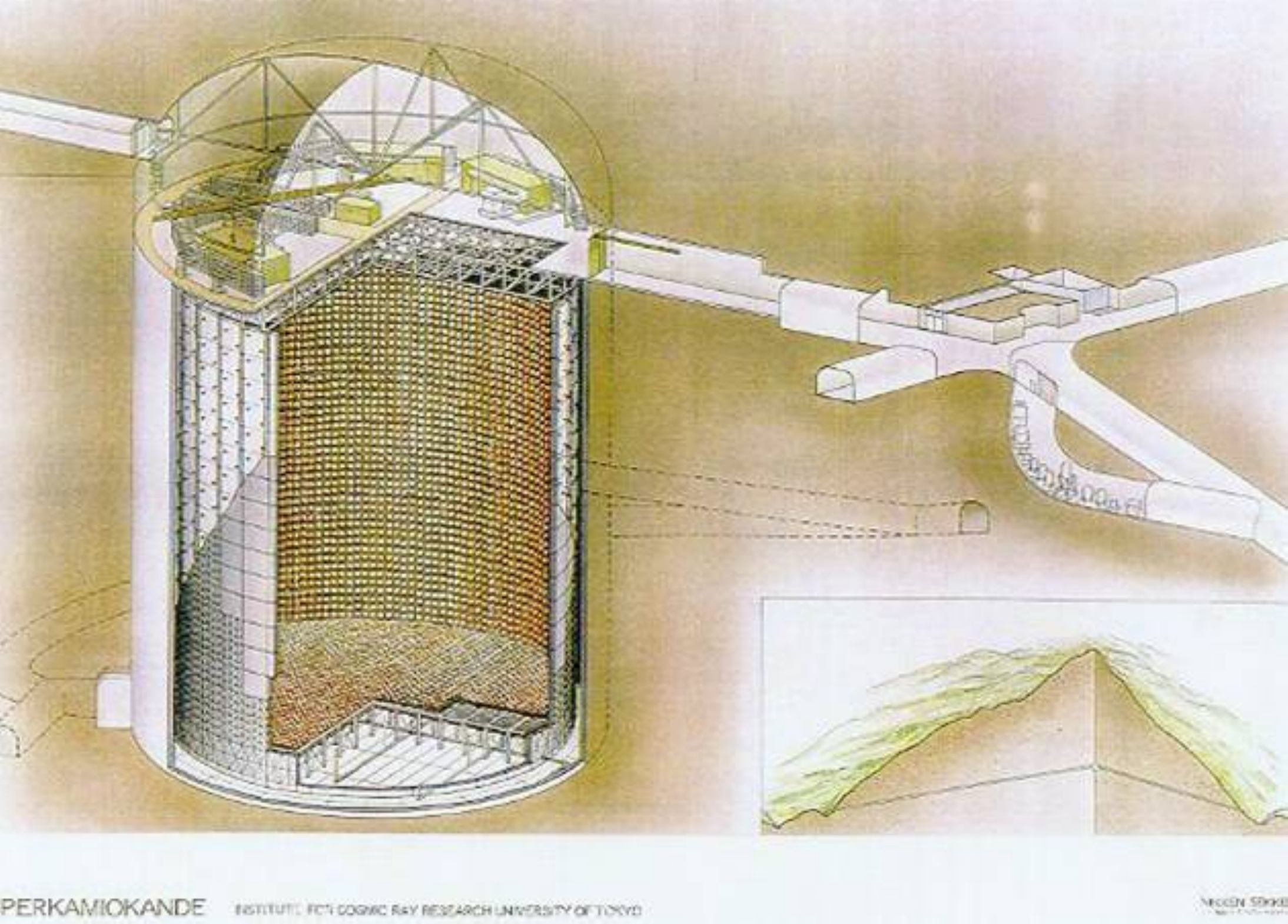}
\caption{Schematic view of the Super-Kamiokande detector.}
\label{SKdet}
\end{center}
\end{figure}

The Super-Kamiokande (SK) detector is a 50,000-ton water Cherenkov detector
located 1000 meters underground in the Kamioka Mine in Japan.
Yoji Totsuka led the detector construction and initial analyses of SK data.
The detector has an inner active volume (32,000 tons) viewed by 11,146 
20-inch diameter photomultipliers (PMTs).
The fiducial volume for the solar 
neutrino measurement is 22,500 tons, defined by the volume more than 2 m from
the surface of the PMTs. 
A schematic view of the SK detector is shown in Fig.\ref{SKdet}.
SK has measured \b8 neutrinos using neutrino--electron scattering in the
same manner as Kamiokande.
The main difference between Super-Kamiokande and Kamiokande is the 30 times
larger fiducial volume and increased fraction of photo-sensitive 
coverage by a factor of 2, which enabled the energy threshold to be lowered below
5 MeV.

To measure the \b8 solar neutrino flux energy spectrum with
high precision, special care was taken in SK.
The absolute energy of the detector was calibrated using an electron linear
accelerator (LINAC)\cite{LINACPAP} installed at the detector site.
The LINAC system could generate mono-energetic electrons
and inject them at various positions in the detector.
The LINAC system provided a very precise energy scale calibration, but it is 
only accurate for
vertical downward-going events. To calibrate the angular dependence of
the energy scale, a $^{16}$N radioactive source\cite{SKN16} was used.
$^{16}$N atoms are produced by fast neutron capture by oxygen nuclei in water.
Neutrons were generated by a commercially built deuteron--tritium generator   
which produced 10$^6$ 14.2-MeV neutrons per pulse.
The main decay mode of $^{16}$N is an electron
with a 4.3-MeV maximum energy coincident with a 6.1-MeV gamma ray.
A setup of the DT generator was deployed in the SK tank, and it is raised
by about 2 m after it emits neutrons in order to avoid shadowing 
of the Cherenkov light.
Because of the precise energy calibration by the LINAC and DT systems, 
the absolute energy scale of the SK detector was calibrated with an
accuracy of 0.64\% (rms) in the first phase of the SK detector (SK-I) and
was improved to 0.53 \% in the fourth phase. 

\begin{figure}
\begin{center}
\includegraphics[width=12cm]{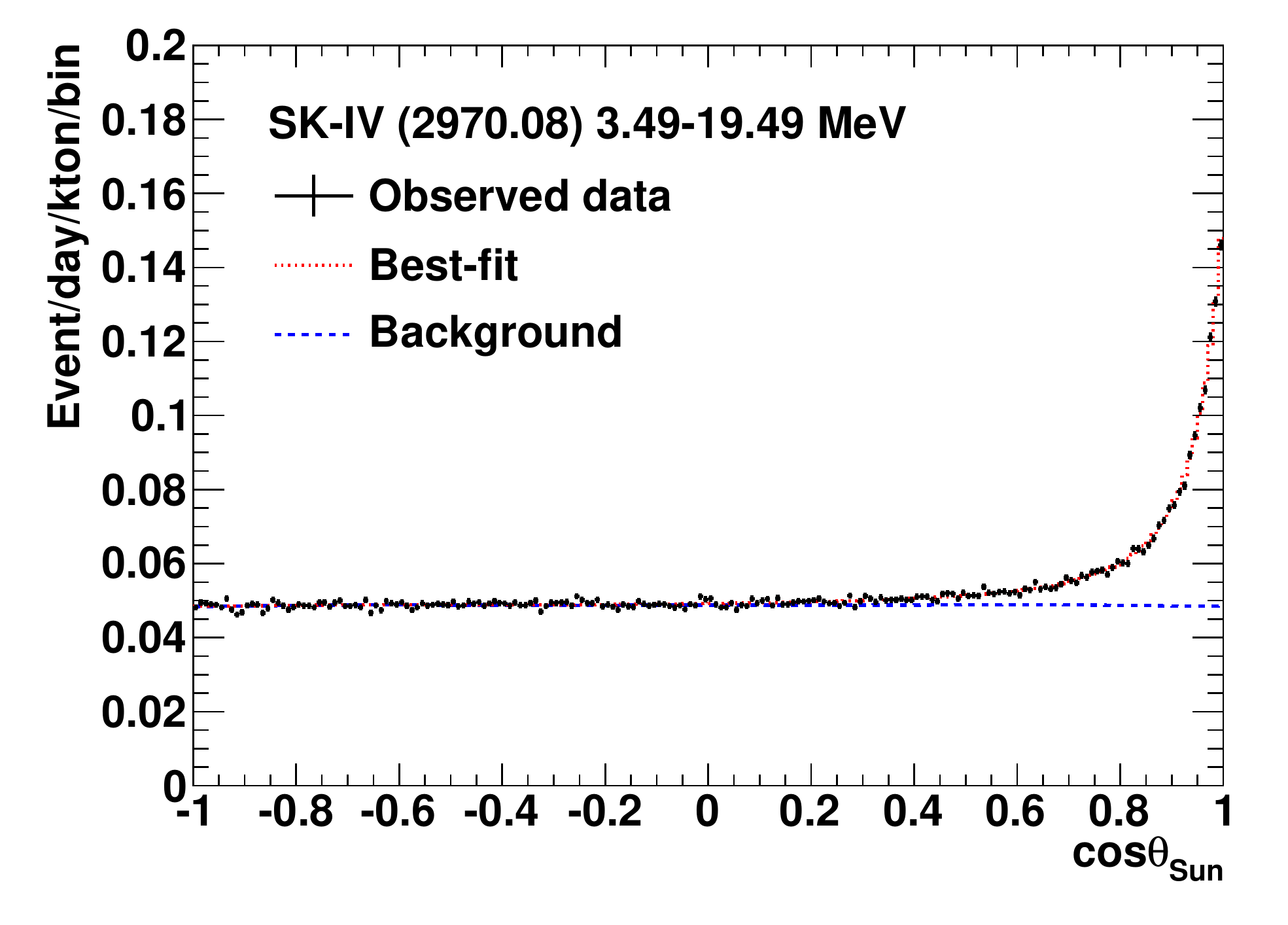}
\caption{Angular distribution with respect to the direction of the sun in
Super-Kamiokande phase IV\cite{SOLSK2020}.}
\label{solskang4}
\end{center}
\end{figure}

SK-I ran for 1496 live-days from May 1996 to July
2001\cite{SOLSK2001, SOLSKFULL, SOLSK2002}.
In November 2001, a chain reaction implosion
destroyed more than half of the PMTs, such that
the second phase (SK-II) ran for 791 live-days
from December 2002 to October 2005, using
5,182 ID PMTs with 19\% photocathode coverage\cite{SOLSK2008}.
Since SK-II the ID PMTs have been covered in fiber-reinforced
plastic cases with an acrylic window to prevent
similar accidents.
The detector was fully reconstructed from October 2005 to July 2006 and
the third phase (SK-III) ran for 548 live-days from October 
2006 to August 2008\cite{SOLSK2010}.
The readout electronics were replaced in September 2008 and the fourth
phase (SK-IV) ran for 2970 live-days until May 2018\cite{SOLSK2020}.
The SK tank was refurbished to make it leak-tight for a
future upgrade (as explained later) from June 2018 to January 2019.
The fifth phase (SK-V) ran from February 2019 through June 2020 with
pure water.
13 tons of Gd$_2$(SO$_4$)$_3\cdot$H$_2$O was loaded into the tank water
to make a 0.01 \%wt Gd solution from July to
August in 2020 and the sixth phase (SK-VI) is running with a
neutron tagging capability.
Solar neutrino data analyses were performed for all SK phases and
were completed by SK-IV.
Figure \ref{solskang4} shows the angular distribution of solar neutrino
candidates with respect to the direction of the sun from SK-IV.
Solar neutrino events are clearly seen above the flat background distribution.
Table \ref{sk-flux-table} shows live-times, energy thresholds, measured
fluxes, ratios compared with a Monte Carlo simulation (in which \b8 flux of
5.25$\times 10^6$/cm$^2$/s is assumed) and the numbers of
extracted signals for each SK phase, and combined results.
\begin{table}
\begin{tabular}{|c|c|c|c|c|c|}
\hline
Phase & Livetime & Threshold & Measured Flux & DATA/MC & \# of signals \\
      & (days)   & (MeV) & ($10^6$/cm$^2$/s) & & ($\pm$stat. only) \\
\hline \hline
SK-I & 1496 & 4.5 & 2.38$\pm 0.02 \pm 0.08$ & 0.453$\pm 0.005 ^{+0.016}_{-0.014} $ & 22443$^{+227}_{-225}$ \\
\hline
SK-II & 791 & 6.5 & 2.41$\pm 0.05 ^{+0.16}_{-0.15}$ & 0.459$\pm 0.010 \pm 0.030$ & 7210$^{+153}_{-151}$ \\
\hline
SK-III & 548 & 4.0 & 2.40$\pm 0.04 \pm 0.05$ & 0.458$\pm 0.007 \pm 0.010$ & 8148$ \pm 133$ \\
\hline
SK-IV & 2970 & 3.5 & 2.33$\pm 0.01 \pm 0.03$ & 0.443$\pm 0.003 \pm 0.006$ & 63890$ ^{+381}_{-379}$\\
\hline \hline
Combined & 5805 & - & 2.35$\pm 0.01 \pm 0.04$ & 0.447$\pm 0.002 \pm 0.008$ & More than\\
 & & & & & 100k events \\
\hline
\end{tabular}
\caption{Summary of flux measurements at Super-Kamiokande.}
\label{sk-flux-table} 
\end{table}
The energy threshold of the data analysis was 4.5 MeV (kinetic energy)
in SK-I but was lowered
to 3.5 MeV in SK-IV. That was possible because the radon background in the tank water
was reduced by a stable laminar flow in the tank given by a precise
temperature control.

\begin{figure}
\begin{center}
  \includegraphics[width=12cm]{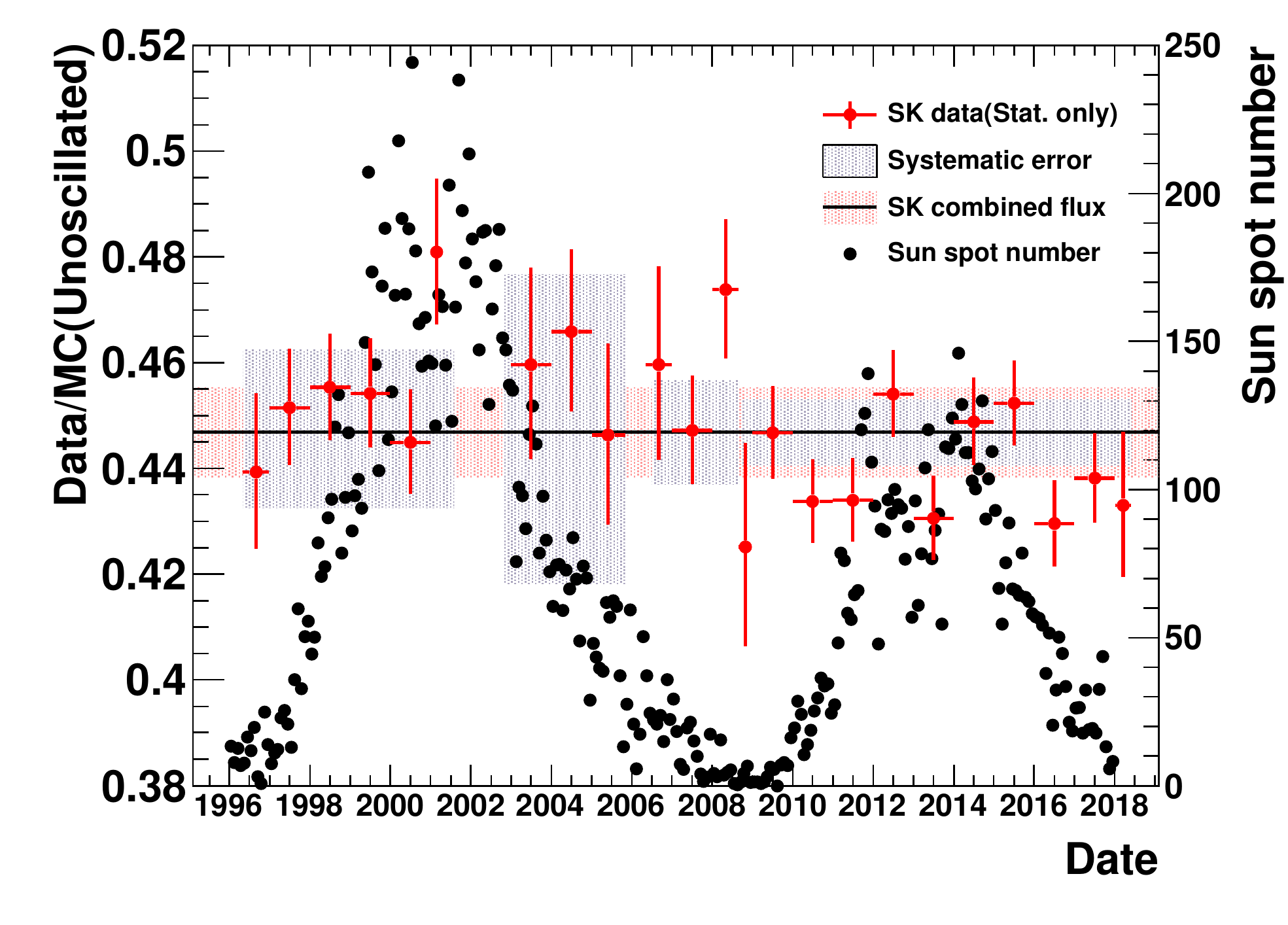}
  \caption{Yearly plot of the \b8 solar neutrino flux measured by SK, shown as red points with error bars.
    The black dots show the number of sunspots, from \cite{SUNSPOT}.}
  \label{skyearlyplot}
\end{center}
\end{figure}

In addition, the systematic error was reduced by more precise calibrations and
detailed studies of the efficiencies of various cuts.
When SK started data collection in 1996, the solar activity was near the minimum.
The SK data covered almost two solar cycles until the end of SK-IV.
Figure \ref{skyearlyplot} shows the measured yearly fluxes compared with the
solar activity. The measured fluxes are consistent with a flat distribution within
statistical and systematic errors and no correlation with the solar activity
was observed.

\begin{figure}
\begin{center}
\includegraphics[width=12cm]{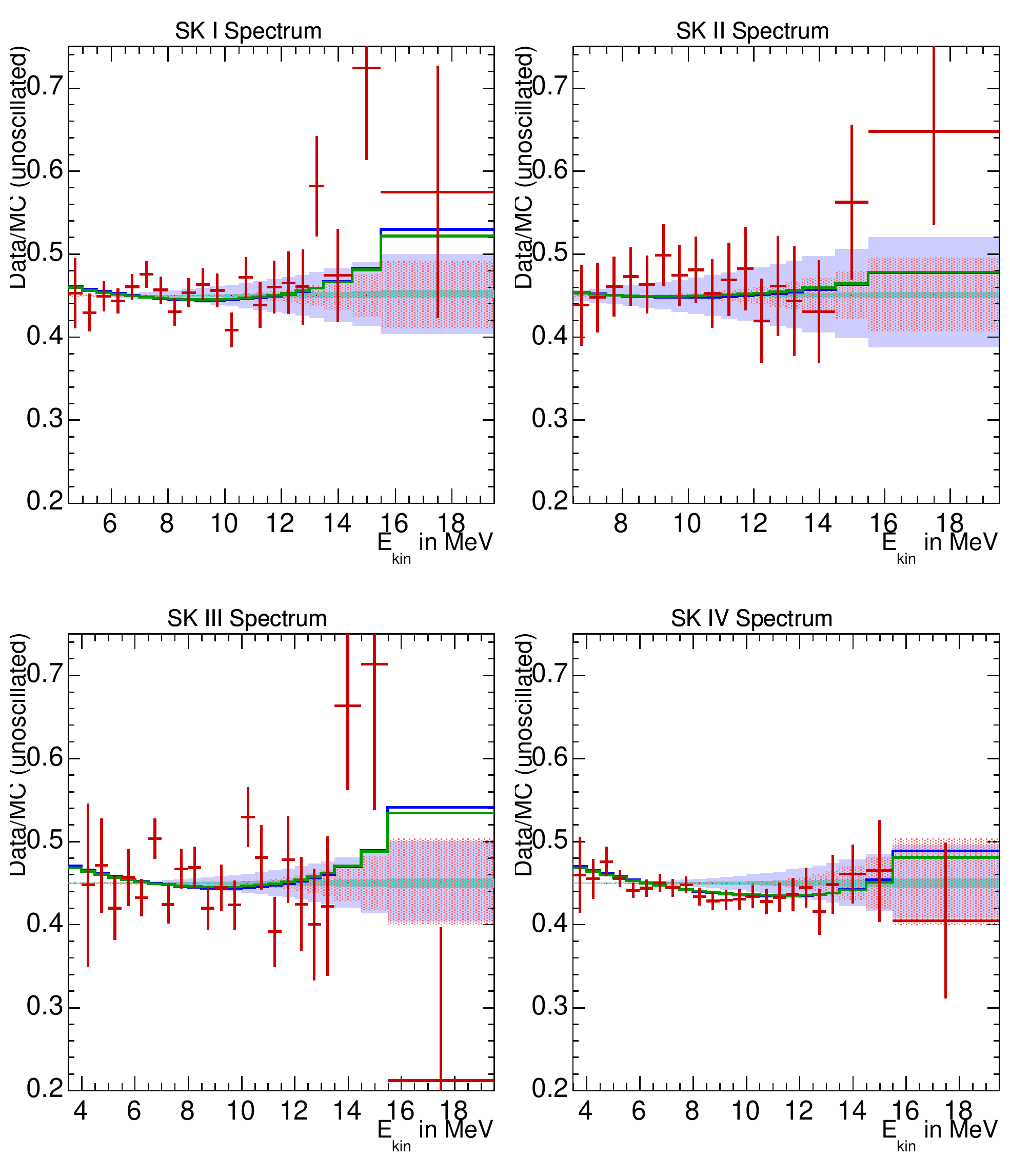}
\caption{Energy spectrum of the solar neutrino signal observed in each SK phase.
The vertical axis is normalized by the spectrum without neutrino oscillations
with a flux of 5.25$\times 10^6$/cm$^2$/s. The blue and green histograms show
the expectations with neutrino oscillations with parameters from the KamLAND
reactor measurement\cite{KAMLAND} and solar global analysis, respectively.
The bands around the horizontal line show $\pm 1 \sigma$ energy-correlated
systematic errors.}
\label{SKspect}
\end{center}
\end{figure}

The energy spectrum shape is important for the discussion of
neutrino oscillations.
Figure \ref{SKspect} shows energy spectra of solar neutrino signals
observed in each SK phase compared with the expected spectrum without
neutrino oscillations.
A \b8 flux of 5.25$\times 10^6$/cm$^2$/s is assumed in this figure, based on the SNO neutral current measurement, consistent with the SSM
predictions.
The observed energy spectrum shapes in SK-I, II and III are consistent with a flat oscillation
probability.
The SK-IV spectrum shape disfavors a flat probability with a significance of $\sim 1 \sigma$, which is consistent with the shape expected from the oscillation parameters obtained
by the KamLAND reactor measurement\cite{KAMLAND}. 

\begin{figure}
\begin{center}
  \includegraphics[width=12cm]{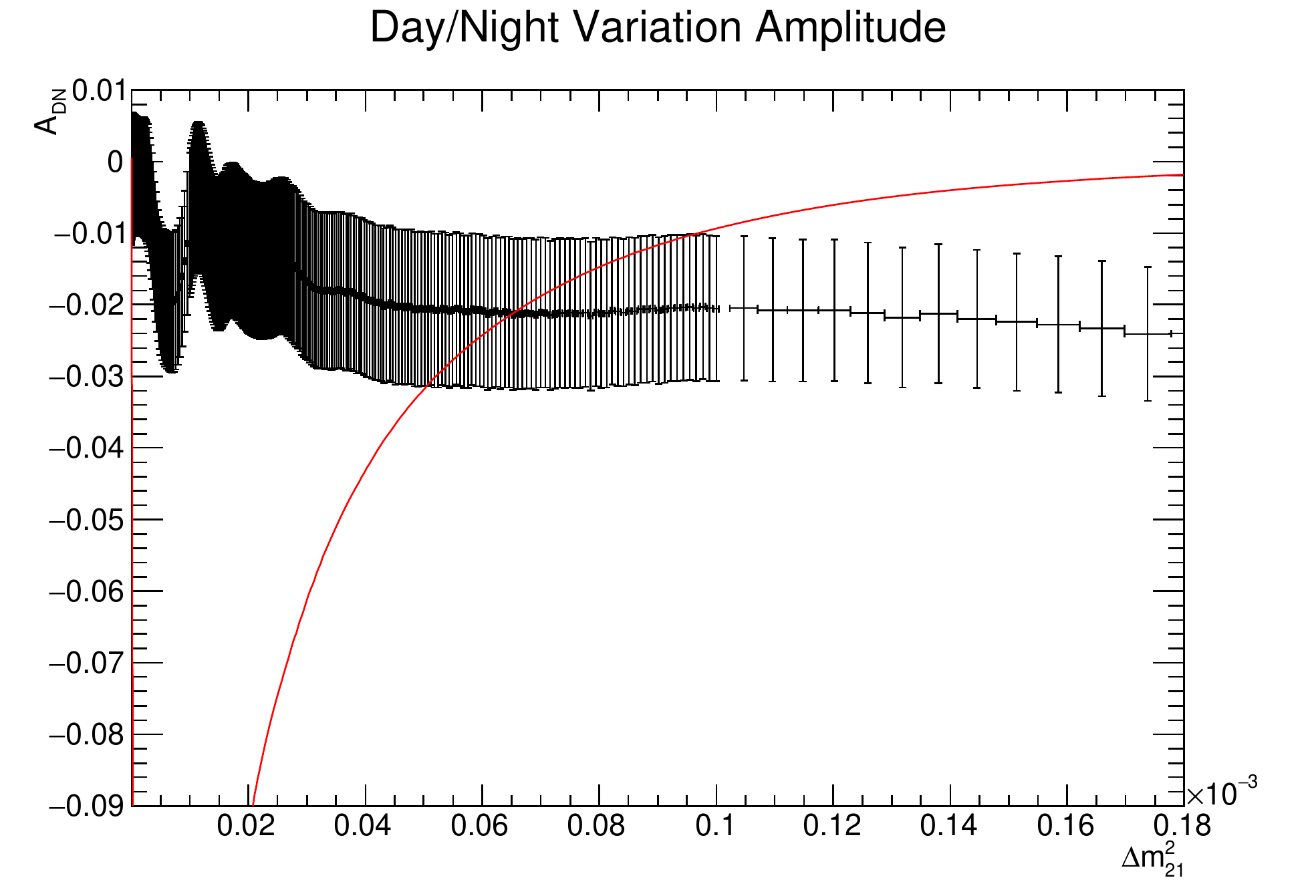}
  \caption{Expected asymmetry from matter oscillation
    in the earth as a function of $\Delta m^2_{21}$ (red curve) and SK-IV data with its statistical error (black data points).}
    \label{daynightasym}
\end{center}
\end{figure}

The day/night flux difference was evaluated by an asymmetry parameter
($A_{DN}$) defined as 
$\frac{(day-night)}{\frac{1}{2}(day+night)}$.
The asymmetry parameters measured by SK-I\cite{SOLSK2002}, SK-II\cite{SOLSK2008},
SK-III\cite{SOLSK2010} and SK-IV\cite{SOLSK2020} were
\begin{eqnarray}
	A^{SK-I}_{DN} &=& -0.021 \pm 0.020\mbox{(stat.)} 
	 ^{+0.013}_{-0.012} \mbox{(sys.)} \nonumber \\
	A^{SK-II}_{DN} &=& -0.063 \pm 0.042\mbox{(stat.)} 
	 \pm 0.037 \mbox{(sys.)} \nonumber \\
	A^{SK-III}_{DN} &=& -0.056 \pm 0.031\mbox{(stat.)} 
	 \pm 0.013 \mbox{(sys.)} \nonumber \\
	A^{SK-IV}_{DN} &=& -0.021 \pm 0.011\mbox{(stat.)} 
	 \hspace{1cm}\mbox{(preliminary)} \nonumber 
\end{eqnarray}
Since $A_{DN}$ depends on $\dms$, as shown by the red curve in
Fig.\ref{daynightasym},  we can measure $\dms$ by $A_{DN}$.
This is a unique method for measuring $\dms$ using solar neutrinos and
is likely to be the only method for neutrinos.
(Note that $\dms$ can be measured for anti-neutrinos using
reactor neutrinos.)
To demonstrate the accuracy of the $\dms$ measurement,
$A_{DN}$ for SK-IV is shown in Fig.\ref{daynightasym} by
black data points with statistical errors.

\section{SNO}\label{SECSNO}

The Sudbury Neutrino Observatory (SNO) detector was a 1000-ton heavy
water (D$_2$O) Cherenkov detector located 2090 meters underground in
the Creighton Mine near Sudbury, Canada.  
It used 9456 8-inch diameter PMTs to view heavy water contained in an acrylic vessel.
The SNO detector could measure the
$\nu_e$ flux from $^8$B neutrinos and the flux of all active
neutrino flavors through the following interactions:
\[\begin{array}{lcll}
    \nu_{e}+d & \rightarrow & p+p+e^{-} & \hspace{0.05in} \mbox{(charged current (CC))}\\ 
    \nu_{x}+d & \rightarrow & p+n+\nu_{x} & \hspace{0.05in} \mbox{(neutral current (NC))} \\
    \nu_{x}+e^{-} & \rightarrow &  \nu_{x}+e^{-} & \hspace{0.05in} \mbox{(electron scattering (ES))} \\
\end{array}\]
where $\nu_x$ is any of $\nu_e$, $\nu_\mu$ or $\nu_\tau$.
The first phase of SNO data collection (SNO-I) was undertaken using a pure D$_2$O
target over 306 days from November 1999 to May 2001\cite{SOLSNO1}.
Free neutrons from the NC interaction were thermalized and 6.25-MeV 
$\gamma$ rays were emitted following their capture by deuterons.
The capture efficiency was about 30\%.
The measured fluxes were
\begin{eqnarray}
    \phi(^8{\rm B})^{CC}_{SNO-I} &=& 
    1.76\pm 0.05\mbox{(stat.)}\pm0.09 \mbox{(sys.)} 
        \times 10^6 /cm^2/s \nonumber \\ 
    \phi(^8{\rm B})^{ES}_{SNO-I} &=& 
    2.39^{+0.24}_{-0.23} \mbox{(stat.)}\pm 0.12 \mbox{(sys.)}. 
        \times 10^6 /cm^2/s \nonumber \\
    \phi(^8{\rm B})^{NC}_{SNO-I} &=& 5.09^{+0.44}_{-0.43} \mbox{(stat.)}
        ^{+0.46}_{-0.43}\mbox{(sys.)} \times 10^6 /cm^2/s~. \nonumber 
\end{eqnarray}
In the second phase of the SNO experiment (SNO-II), 2 tons of NaCl were
added to the D$_2$O target to enhance the detection efficiency
of the NC channel\cite{SOLSNO2}.  Thermalized neutrons were captured by 
$^{35}$Cl
nuclei, resulting in the emission of a $\gamma$-ray cascade with a
total energy of 8.6~MeV.
The CC and NC signals were statistically separated using 
the isotropy of the Cherenkov light pattern and each event's angle to the sun.
The forward peaked signal was due to ES and the backward distribution was
due to CC interactions. 
NC events were isotropic with respect to the solar direction.
The measured fluxes in SNO-II were
\begin{eqnarray}
	\phi(^8{\rm B})^{CC}_{SNO-II} &=& 1.68 \pm 0.06\mbox{(stat.)} 
	    ^{+0.08}_{-0.09}\mbox{(sys.)} \times 10^6 /cm^2/s \nonumber \\
	\phi(^8{\rm B})^{ES}_{SNO-II} &=& 2.35 \pm 0.22\mbox{(stat.)} 
	    ^{+0.15}_{-0.15}\mbox{(sys.)} \times 10^6 /cm^2/s \nonumber \\
	\phi(^8{\rm B})^{NC}_{SNO-II} &=& 4.94 \pm 0.21\mbox{(stat.)} 
	    ^{+0.38}_{-0.34}\mbox{(sys.)} \times 10^6 /cm^2/s . \nonumber 
\end{eqnarray}

In the third phase of SNO (SNO-III), $^3$He proportional counters were deployed
in the heavy water and the NC events were measured independently\cite{SOLSNO3}.
The NC flux measured by SNO-III was
\begin{eqnarray}
	\phi(^8{\rm B})^{NC}_{SNO-III} &=& 5.54 ^{+0.33}_{-0.31}\mbox{(stat.)} 
	    ^{+0.36}_{-0.34}\mbox{(sys.)} \times 10^6 /cm^2/s . \nonumber 
\end{eqnarray}

The SNO group performed a combined analysis\cite{SOLSNOCOMB} of all three phases
and the obtained result was
\begin{eqnarray}
	\phi(^8{\rm B})^{NC}_{SNO~combined} &=& 5.25 \pm 0.16 \mbox{(stat.)} 
	^{+0.11}_{-0.13}\mbox{(sys.)} \times 10^6 /cm^2/s . \nonumber
\end{eqnarray}
Compared with the \b8 flux value in Table \ref{sol-ssmtable}, the observed
\b8 flux with NC by SNO agrees well with the prediction within the errors.
Combining the statistical and systematic errors, the total error of the flux
observed by SNO is 3.8\%, which is better than that of the SSM prediction
by a factor of 3.
Therefore, the SNO value was used to make plots of SK in the previous section and
will be used in the discussion of the global oscillation analysis in
Section \ref{SECOSC}.

\section{Borexino} 

Borexino is a liquid scintillator detector with an active mass of 278 tons
of pseudocumene located in the Gran Sasso Laboratory.
Scintillation light is detected via 2212 8-inch PMTs uniformly distributed 
on the inner surface of the detector.
Because of the high light yield of a liquid scintillator compared with 
Cherenkov light, Borexino is sensitive to sub-MeV solar neutrinos.
The first $^7$Be solar neutrino measurement was reported 
in \cite{Borexino-1st}, based on 192 days of data taken from May 2007 to April
2008.
The 0.862-MeV monoenergetic $^7$Be neutrinos were detected by neutrino--electron
scattering. 
The energy spectrum of observed events was deconvoluted using the expected shape
of the recoil electrons and possible background sources.
Thus, the extracted $^7$Be neutrino event rate was
\begin{eqnarray}
49 \pm 3\mbox{(stat.)} \pm 4 \mbox{(sys.)}~ {\rm counts} / ({\rm day} \cdot 100 {\rm ton)}, \nonumber 
\end{eqnarray}
while the expected event rate from the SSM 
was $74 \pm 4 ~ {\rm counts} / ({\rm day} \cdot 100 {\rm ton})$.
The deficit of the observed $^7$Be neutrino event rate is consistent with the
expectation based on neutrino oscillations.
Borexino also succeeded in observing $pep$, $pp$ and CNO solar neutrinos, reported in \cite{Borexino-pep}, \cite{Borexino-pp}  and \cite{Borexino-CNO},
respectively.
The Borexino group corrected the effect of neutrino oscillations and obtained
fluxes of $pp$, $^7$Be, $pep$ and CNO neutrinos from the second phase of
Borexino as\cite{Borexino2,Borexino-CNO}
\begin{eqnarray}
	\phi(pp)^{Osc. corrected}_{Borexino} &=& 6.1 \pm 0.5\mbox{(stat.)} 
	    ^{+0.3}_{-0.5}\mbox{(sys.)} \times 10^{10} /cm^2/s \nonumber \\
	\phi(^7{\rm Be})^{Osc. corrected}_{Borexino} &=& 4.99 \pm 0.11\mbox{(stat.)} 
	    ^{+0.06}_{-0.08}\mbox{(sys.)} \times 10^9 /cm^2/s \nonumber \\
	\phi(pep(HZ/LZ))^{Osc. corrected}_{Borexino} &=& 1.27/1.39 \pm 0.19\mbox{(stat.)} 
	    ^{+0.08}_{-0.12/-0.13}\mbox{(sys.)} \times 10^8 /cm^2/s \nonumber \\
	\phi({\rm CNO})^{Osc. corrected}_{Borexino} &=& 7.0 ^{+3.0}_{-2.0}\mbox{(sys.)} \times 10^8 /cm^2/s \nonumber
\end{eqnarray}
These fluxes are compared with the SSM predictions in Table \ref{sol-ssmtable}
and are consistent with each other within the experimental errors.

\section{Solar Neutrino Oscillations}\label{SECOSC}

Precise solar neutrino measurements have shown that the solar neutrino problem is
due to neutrino oscillations.
In this section, a brief history of and the latest results for solar neutrino oscillations
are discussed.

\subsection{Evidence of solar neutrino oscillations}

\begin{figure}
\begin{center}
\includegraphics[width=12cm]{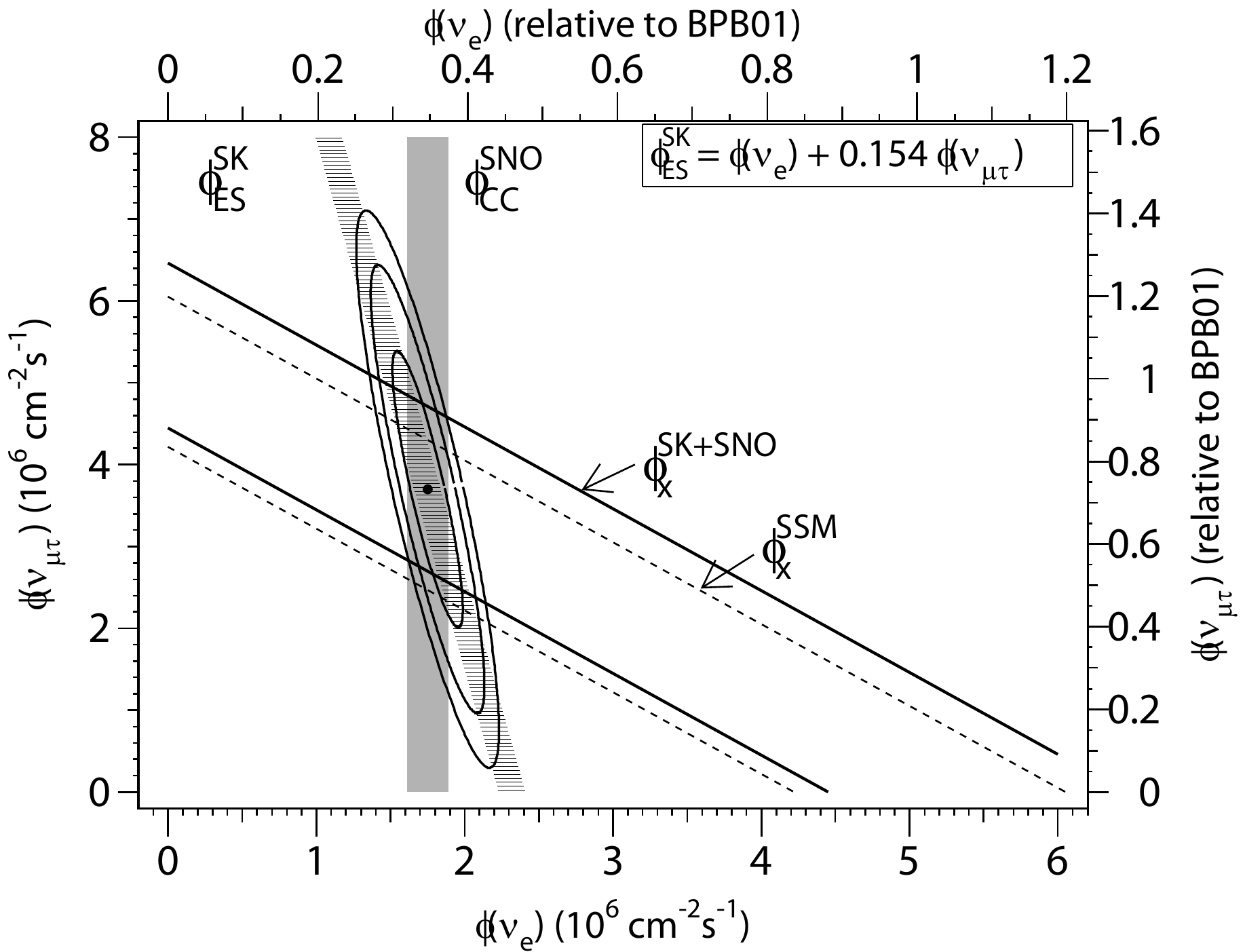}
\caption{$\nue$ flux and $\nu_{\mu/\tau}$ flux contour obtained from the
  results of SK and SNO in 2001\cite{SNOfirstCC}.
  Each band shows the $\pm 1 \sigma$ range of
  each measurement.}
\label{SKSNO2001}
\end{center}
\end{figure}

\begin{figure}
\begin{center}
\includegraphics[width=12cm]{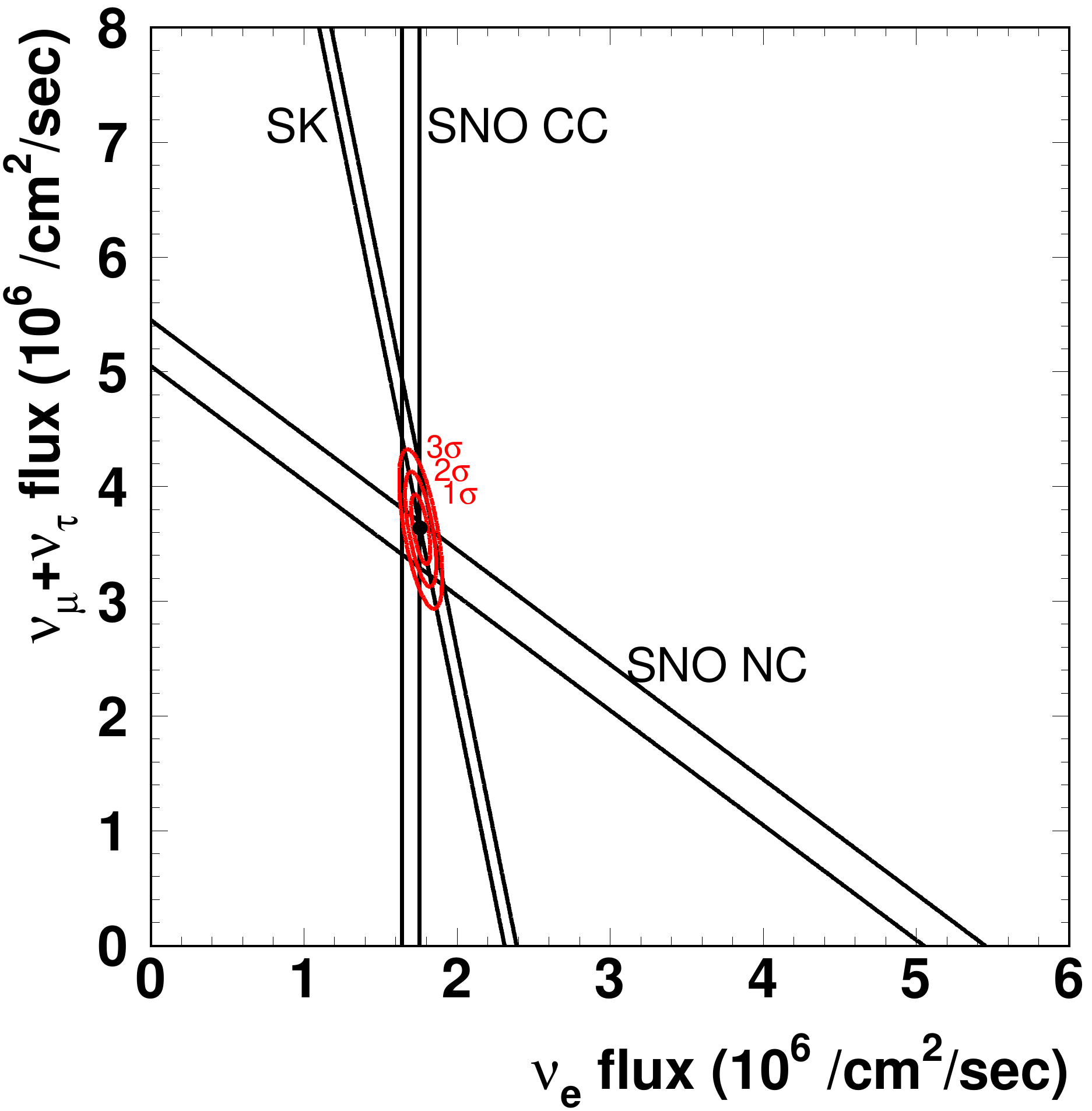}
\caption{Latest $\nue$ flux and $\nu_{\mu/\tau}$ flux contour obtained from the
  results of SK\cite{SOLSK2020} and SNO\cite{SOLSNO1,SOLSNO2,SOLSNO3,SOLSNOCOMB}.
  Each band shows the $\pm 1 \sigma$ range for each measurement.}
\label{SKSNO2022}
\end{center}
\end{figure}

In March 2001, the SK collaboration released results of the precise measurement of
the \b8 flux using 1258 days of SK data\cite{SOLSK2001}.
SK measured \b8 solar neutrinos using neutrino--electron scattering (ES), in which
$\numu$ and $\nutau$ contribute in addition to $\nue$ because the
cross section of $\nu_{\nu/\tau}e^-$ scattering is about 1/($6\sim7$) of that
for $\nu_ee^-$ scattering.
In June 2001, the SNO collaboration released the first result of the charged current (CC),
i.e. $\nue$, measurement,
including the plot in Fig.\ref{SKSNO2001}, which demonstrated that
$\nu_{\mu/\tau}$ exist with more than 3$\sigma$ significance\cite{SNOfirstCC}.
That was the first direct evidence for solar neutrino oscillations.
The SNO collaboration published NC current measurements, the results of which
were described above in section \ref{SECSNO}.
The SK collaboration provided further data for analysis and improved the accuracy of the ES measurement.
The plot in Fig.\ref{SKSNO2022} uses the latest data from SK and SNO, including NC data,
showing that the significance of
$\nu_{\mu/\tau}$ had been greatly improved.
The measurements of SK and SNO are consistent with each other.

\subsection{Determination of oscillation parameters}

In the three flavor neutrino framework, the relation between the
mass and interaction eigenstates is described by
\begin{eqnarray}
\left(
\begin{array}{c}
\nue \\
\numu \\
\nutau \\
\end{array}
\right) &=&
\left(
\begin{array}{ccc}
U_{e1} & U_{e2} & U_{e3} \\
U_{\mu 1} & U_{\mu 2} & U_{\mu 3} \\
U_{\tau 1} & U_{\tau 2} & U_{\tau 3} \\
\end{array}\right)
\left(
\begin{array}{c}
\nu_1 \\
\nu_2 \\
\nu_3 \\
\end{array}
\right) 
\nonumber
\end{eqnarray}
The unitary matrix $U$ is the Pontecorvo--Maki--Nakagawa--Sakata (PMNS)
matrix and can be decomposed into three angles and a phase:
\begin{eqnarray}
    U&=&
    \begin{pmatrix}
    1&0&0 \cr
    0& {c_{23}} & {s_{23}} \cr
    0& -{s_{23}}& {c_{23}}\cr
\end{pmatrix}    
\begin{pmatrix}
    {c_{13}} & 0 & {s_{13}}e^{-i {\delta}}\cr
    0&1&0\cr 
    -{ s_{13}}e^{i {\delta}} & 0  & {c_{13}}\cr
\end{pmatrix}    
\begin{pmatrix}
    c_{21} & {s_{12}}&0\cr
    -{s_{12}}& {c_{12}}&0\cr
    0&0&1\cr
    \end{pmatrix}\nonumber \\
&=& 
\begin{pmatrix}
c_{12}c_{13} & s_{12}c_{13} &  s_{13}e^{-i\delta} \cr
-s_{12}c_{23}-c_{12}s_{23}s_{13}e^{i\delta} & 
c_{12}c_{23}-s_{12}s_{23}s_{13}e^{i\delta} & s_{23}c_{13} \cr
s_{12}s_{23}-c_{12}c_{23}s_{13}e^{i\delta} & 
-c_{12}s_{23}-s_{12}c_{23}s_{13}e^{i\delta} & c_{23}c_{13} \cr
\end{pmatrix} ~~, \nonumber
\end{eqnarray}
where $c_{ij} \equiv \cos\theta_{ij}$ and $s_{ij} \equiv \sin\theta_{ij}$.
In addition, the mass squared differences ($\Delta m^2_{ij}(=m^2_i-m^2_j)$) 
$\Delta m^2_{21}$ and $\Delta m^2_{32}$ affect the oscillations.
In the solar neutrino oscillation analysis, the oscillation
probability can be calculated 
using three parameters, $\theta_{12}$, $\theta_{13}$ and 
$\Delta m^2_{21}$, because 
$|\Delta m^2_{21}| \ll |\Delta m^2_{32}|$. 

\begin{figure}
\begin{center}
\includegraphics[width=12cm]{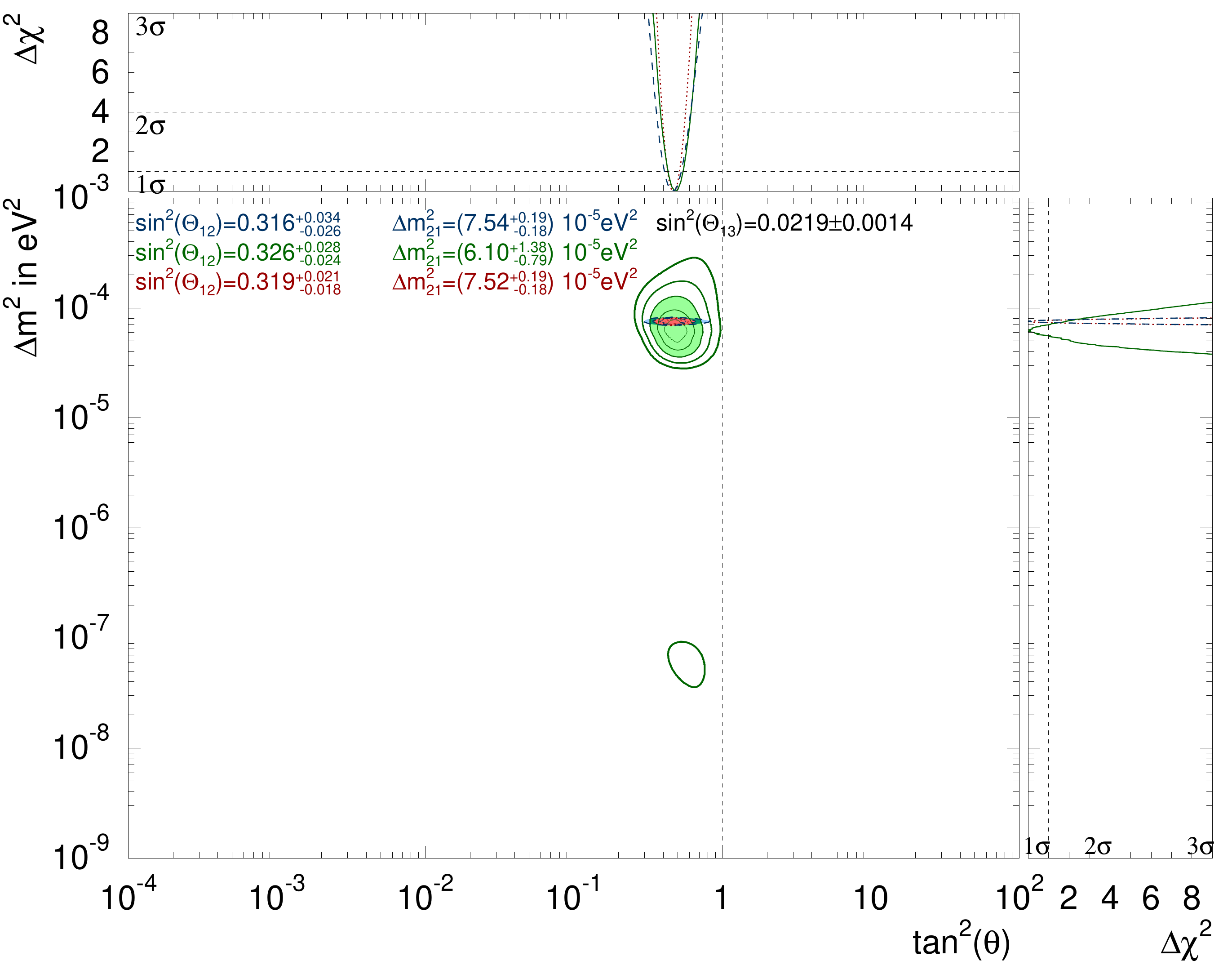}
\caption{Oscillation parameters obtained from SK data\cite{SOLSK2020} compared with
  the KamLAND reactor data\cite{KAMLAND}.
  The green contours show the allowed region of $\theta_{12}$ and $\Delta m^2_{21}$
  obtained from the SK data with the \b8 flux constrained by the
  SNO NC value((5.25$\pm 0.20)\times 10^6$/cm$^2$/s).
  The lines show 1, 2, 3, 4 and 5$\sigma$ significance from inside to outside.
  The light-blue filled area shows the KamLAND reactor contour with 3$\sigma$ and
  the brown filled area shows the SK and KamLAND combined result with 3$\sigma$.}
\label{skcont-log}
\end{center}
\end{figure}

\begin{figure}
\begin{center}
\includegraphics[width=12cm]{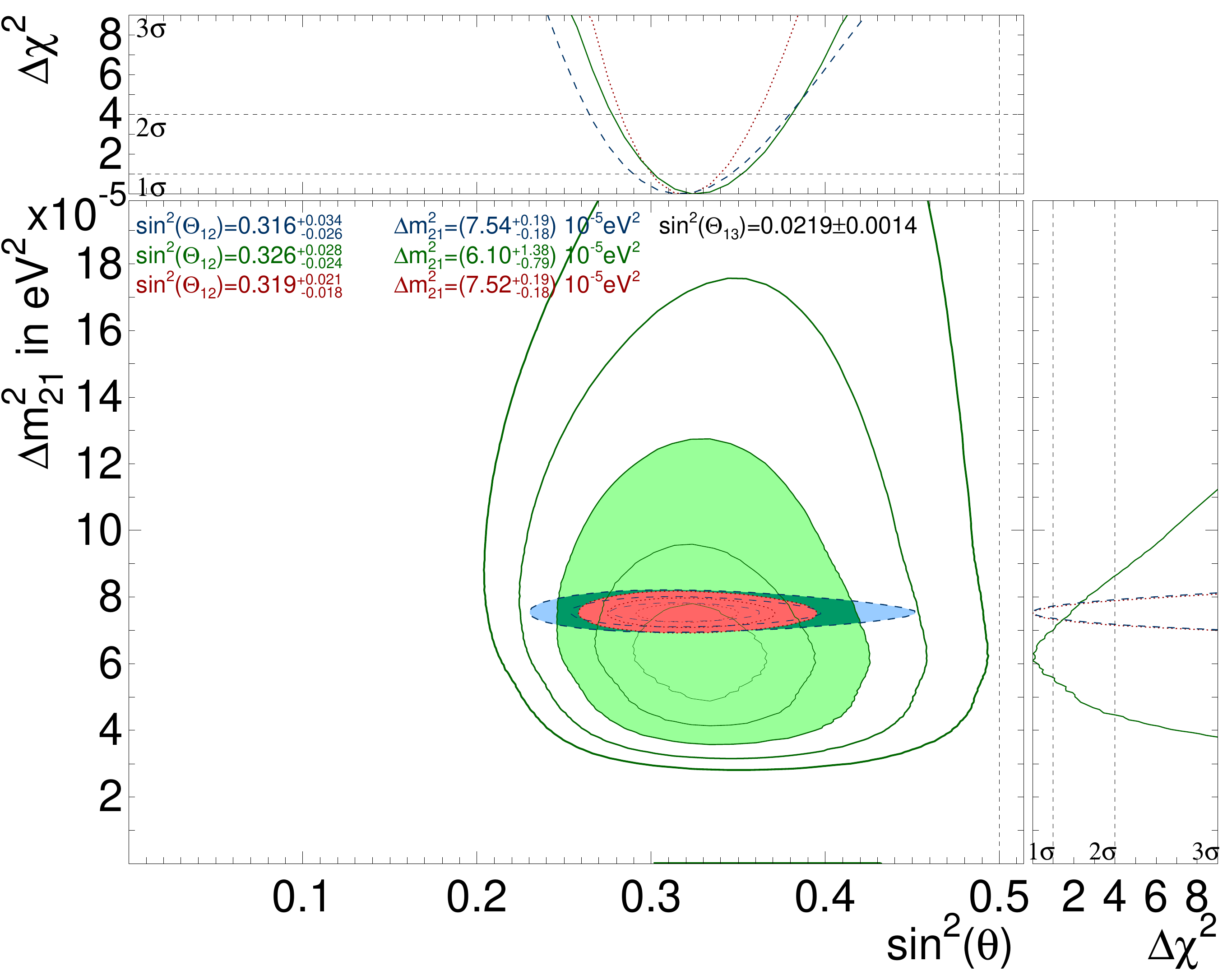}
\caption{As for Fig.\ref{skcont-log} but the vertical axis has a linear scale.
  The green, light-blue and brown regions show contours of SK, KamLAND and
  the combination of SK and KamLAND, respectively.}
\label{skcont-linear}
\end{center}
\end{figure}

\begin{figure}
\begin{center}
\includegraphics[width=12cm]{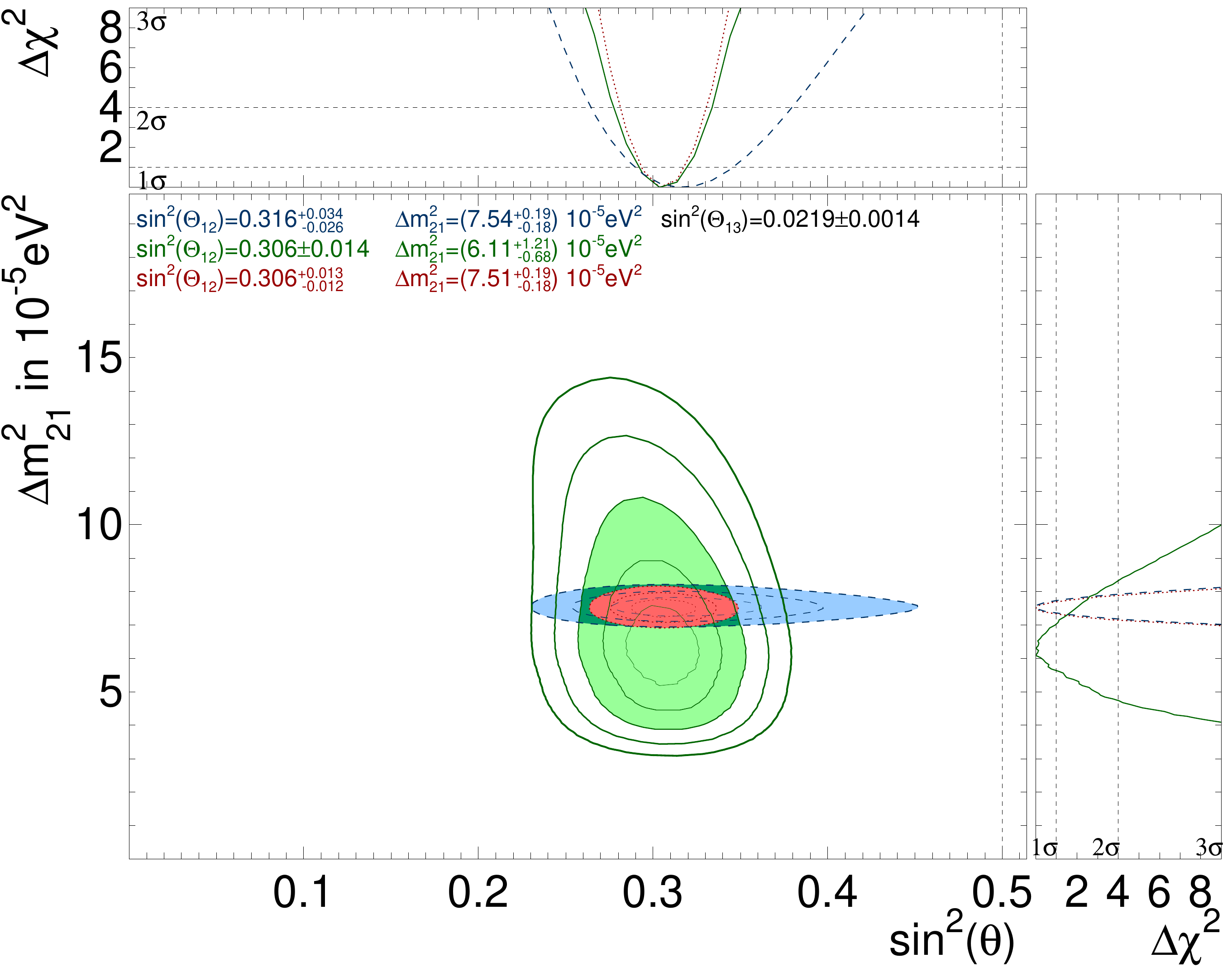}
\caption{Oscillation parameters obtained from SK and SNO data compared with
  the KamLAND reactor data\cite{KAMLAND}.
  The green contours show the allowed region of $\theta_{12}$ and $\Delta m^2_{21}$
  obtained from the SK and SNO data.
  The lines show the 1, 2, 3, 4 and 5$\sigma$ significance from inside to outside.
  The light-blue filled area shows the KamLAND reactor contour with 3$\sigma$, and
  the brown filled area shows the SK, SNO and KamLAND combined contour with 3$\sigma$.}
  \label{sksnocont-linear}
\end{center}
\end{figure}

The green contours in Fig.\ref{skcont-log} show the allowed region of the oscillation
parameters, $\theta_{12}$ and $\Delta m^2_{21}$, obtained from the SK data.
The total flux of \b8 is constrained by the SNO NC flux measurement 
((5.25$\pm 0.20)\times 10^6$/cm$^2$/s).
The $\theta_{13}$ value is constrained to $\sin^2(\theta_{13})=0.0219 \pm 0.0014$ by the short baseline reactor neutrino measurements\cite{REACTOR13}.
The LMA solution is more than 4$\sigma$-level more significant than the other
solutions and the result is consistent with the long baseline reactor
measurement by KamLAND\cite{KAMLAND} shown by the blue contours in the figure.
The same plot on a linear scale is shown in Fig.\ref{skcont-linear}.
Figure \ref{sksnocont-linear} shows the allowed region obtained by
combining the SK and SNO\cite{SOLSNOCOMB} data.

\section{Conclusion and future prospects}\label{SECCONC}

The solar neutrino problem suggested by the Homestake experiment lead to some very
interesting outcomes.
Owing to the insight of Prof. Koshiba, the solar neutrino problem was
confirmed by a water Cherenkov detector, Kamiokande.
The water Cherenkov technique was adopted in Super-Kamikande and SNO and
precise measurements of the \b8 solar neutrino flux were conducted.
Especially, the flux difference between the neutral current and charged
current methods, which are sensitive to all types of neutrinos and only to $\nue$,
respectively, provided smoking gun evidence of neutrino oscillations.
All the results from solar neutrino experiments (Homestake, Kamiokande, SAGE, GALLEX/GNO,
Super-Kamiokande, SNO and Borexino) can be shown to be consistent with the expectations
from the standard solar model with neutrino oscillations.
That is a great accomplishment and it could not have been done without the
insight of Prof. Koshiba.

Is this the end of the story for solar neutrinos?
I do not think so.
The measurements of the \b8 spectrum shape and the day/night difference are not yet
precise enough to allow a proper discussion of neutrino oscillations.
Borexino succeeded in observing CNO neutrinos but without sufficient precision
to discuss the low and high metallicity models.
The most abundant solar neutrino source is $pp$ neutrinos and the SSM
predicts its flux with an accuracy of $\sim$1\%.
However, the experimental results are not yet this precise.
Something interesting might still be hidden in solar neutrinos.

\section*{Acknowledgment}

The author gratefully acknowledges the cooperation of the Super-Kamiokande
collaborators in providing the main contents of the SK results and oscillation
analyses.
The author also wishes to thank the proposers of this special issue (Profs. Mitsuaki
Nozaki and Masashi Yokoyama) for giving me the opportunity to write this article.
The author thank Prof. S. T. Petcov for sending comments.
%

%
%
%

\let\doi\relax


\end{document}